\documentclass[fleqn,usenatbib]{mnras}

\usepackage{newtxtext,newtxmath}
\usepackage{bm}
\usepackage{float}
\usepackage{subfigure}

\usepackage{scalerel,tikz}
\usetikzlibrary{svg.path}
\definecolor{orcidlogocol}{HTML}{A6CE39}
\tikzset{orcidlogo/.pic={
 \fill[orcidlogocol] svg{M256,128c0,70.7-57.3,128-128,128C57.3,256,0,198.7,0,128C0,57.3,57.3,0,128,0C198.7,0,256,57.3,256,128z};
 \fill[white] svg{M86.3,186.2H70.9V79.1h15.4v48.4V186.2z}
 svg{M108.9,79.1h41.6c39.6,0,57,28.3,57,53.6c0,27.5-21.5,53.6-56.8,53.6h-41.8V79.1z M124.3,172.4h24.5c34.9,0,42.9-26.5,42.9-39.7c0-21.5-13.7-39.7-43.7-39.7h-23.7V172.4z}
 svg{M88.7,56.8c0,5.5-4.5,10.1-10.1,10.1c-5.6,0-10.1-4.6-10.1-10.1c0-5.6,4.5-10.1,10.1-10.1C84.2,46.7,88.7,51.3,88.7,56.8z};
}}
\newcommand\orcidicon[1]{\href{https://orcid.org/#1}{\mbox{\scalerel*{
\begin{tikzpicture}[yscale=-1,transform shape]
\pic{orcidlogo};
\end{tikzpicture}
}{|}}}}

\usepackage[T1]{fontenc}

\DeclareRobustCommand{\VAN}[3]{#2}
\let\VANthebibliography\thebibliography
\def\thebibliography{\DeclareRobustCommand{\VAN}[3]{##3}\VANthebibliography}

\usepackage{graphicx}	
\usepackage{amsmath}	
\usepackage{multirow}

\usepackage[normalem]{ulem} 

\newcommand{\sev}[2]{\bgroup\markoverwith{\textcolor{magenta}{\rule[0.5ex]{2pt}{1pt}}}\ULon{#1}\textcolor{magenta}{#2}}

\title[Phase space chevrons as a subhalo detector]{Ironing the folds: The phase space chevrons of a GSE-like merger as a dark matter subhalo detector}

\author[E. Y. Davies et al.]{
Elliot Y. Davies~\orcidicon{0000-0001-5996-4072}$^{1}$\thanks{E-mail: eyd20@cam.ac.uk},
Eugene Vasiliev~\orcidicon{0000-0002-5038-9267}$^{1}$,
Vasily Belokurov~\orcidicon{0000-0002-0038-9584}$^{1,2}$,
N. Wyn Evans~\orcidicon{0000-0002-5981-7360}$^{1}$, Adam M. Dillamore~\orcidicon{0000-0003-0807-5261}$^{1}$
\\
$^{1}$Institute of Astronomy, University of Cambridge, Madingley Road, Cambridge CB3 0HA, UK\\
$^2$Center for Computational Astrophysics, Flatiron Institute, 162 5th Avenue, New York, NY 10010, USA\\
}

\date{Accepted XXX. Received YYY; in original form ZZZ}

\pubyear{2022}

\begin{document}
\label{firstpage}
\pagerange{\pageref{firstpage}--\pageref{lastpage}}
\maketitle

\begin{abstract}
Recent work uncovered features in the phase space of the Milky Way's stellar halo which may be attributed to the last major 
merger. 
When stellar material from a satellite is accreted onto its host, it \textit{phase mixes} and appears finely substructured in phase space. For a high-eccentricity merger, this substructure most clearly manifests as numerous wrapping chevrons in $(v_r, r)$ space, corresponding to stripes in $(E, \theta_r)$ space.
We introduce the idea of using this substructure as an alternative subhalo detector to cold stellar streams.
We simulate an {\it N}-body merger akin to the GSE and assess the impact of subhaloes on these chevrons. We examine how their deformation depends on the mass, pericentre, and number of subhaloes. To quantify the impact of perturbers in our simulations, we utilise the appearance of chevrons in $(E, \theta_r)$ space to introduce a new quantity -- the \textit{ironing} parameter.
We show that: (1) a single flyby of a massive ($\sim 10^{10}$ M$_{\odot}$) subhalo with pericentre comparable to, or within, the shell's apocentre smooths out the substructure, (2) a single flyby of a low mass ($\lesssim 10^8$ M$_{\odot}$) has negligible effect, (3) multiple flybys of subhalos derived from a subhalo mass function between $10^7-10^{10}$ M$_{\odot}$ cause significant damage if deep within the potential, (4) the effects of known perturbers (e.g. Sagittarius) should be detectable and offer constraints on their initial mass. The sensitivity to the populations of subhaloes suggests that we should be able to place an upper limit on the Milky Way's subhalo mass function.
\end{abstract}




\section{Introduction}

The hierarchical nature of galaxy formation rests on the assumption of the cold dark matter (CDM) model of cosmology \citep[e.g.][]{mo2010galaxy}, whereby galaxies are initially seeded from the primordial overdensities of dark matter. These clumps merge together to create large haloes whose accumulation of mass over time further permits the infall of more smaller DM \textit{subhaloes} \citep[][]{white1978core}. In a galaxy, the expected number of subhaloes in a given subhalo mass bin -- the subhalo mass function (SHMF) -- is intrinsically linked to the model of DM that one assumes. Quantifying the SHMF is therefore a crucial aim of many observational experiments; any deviation from the simulation-derived CDM SHMF \citep[e.g.][]{springel2008aquarius} would therefore provide insight into the validity of the assumptions made in such simulations. While many DM (sub) haloes are massive enough to permit the formation of stellar material, many are not. The relationship between the stellar mass $(M_{\star})$ and halo mass $(M_{\rm h})$ of galaxies suggests that the ratio $M_{\star} / M_{\rm h}$ peaks in haloes of mass $ \sim 10^{12}$ M$_{\odot}$ \citep[][]{behroozi2013average, moster2013galactic}, and that there is a relative deficiency of stars in subhaloes with masses at the lower end of the limit, around $M \sim 10^{7}$ M$_{\odot}$ \citep[][]{read2017stellar,jethwa2018upper,read2019abundance,nadler2020milky, kravtsov2022grumpy}. Therefore, it is anticipated that the Milky Way (MW) should be populated by an ensemble of star-less DM clumps, whose detection via impact on stellar substructure provides a method to probe the nature of DM, and the potential of our Galaxy.

The utilisation of stellar halo substructure in the local Galaxy as a probe for the properties of these subhaloes is a relatively modern idea. Specifically, much recent work has developed the concept of using gaps in cold stellar streams as ``detectors'' for DM subhaloes \citep[e.g.][]{ibata2002uncovering, johnston2002how, gaskins2008signatures, carlberg2009star, erkal2015forensics}. Stellar streams are named so because of their extended leading and trailing arms, which approximately trace out the orbit of their progenitor, and provide a cross-section of interaction with DM subhaloes. The coldest stellar streams result from the tidal dissolution of globular clusters (GC). Thanks to the low internal velocity dispersions, GC streams are more sensitive to smaller perturbers \citep[see e.g.][]{Erkal2015prop}.  While GC streams make for effective detectors of lower mass perturbing subhaloes, they do have limitations. Firstly, streams' long, thin arms provide only a small volume coverage and thus a relatively small cross-section for subhalo interactions \citep[see][]{Erkal2016}. 
Moreover, the possibility of a stream being formed outside of the MW in its previous, dwarf galaxy, host -- the cocoon effect -- adds to the uncertainty in the utility of streams as subhalo detectors \citep[][]{malhan2019butterfly}. Lastly, the long timescale of stream formation means that if a stream is perturbed while its progenitor GC is still undergoing dissolution, the signs of an earlier interaction with a subhalo may get airbrushed by the freshly supplied tidal material \citep[see e.g.][]{dillamore2022impact}.

Motivated to combat the above limitations, we look for a detector with a larger cross-section to DM subhalo interactions. In the wake of the {\it Gaia} mission, mapping the \textit{detailed} stellar halo substructure of the local Galaxy in full 6-d is finally becoming possible \citep[][]{Gaia}. This allows the use of the already known source of substructure with a much larger cross-section than a typical stellar stream or even a population of streams: the remnant of our Galaxy's most massive merger, the \textit{Gaia} Sausage / Enceladus (GSE). The GSE is an ancient $(1 < z < 2)$ merger \citep[][]{belokurov2018co, helmi2018merger}, estimated to have deposited as much as 20\% of the inner Galactic halo's dark matter content and as much as 2/3 of its stellar content \citep[][]{fattahi2019origin, dillamore2022merger,naidu2021reconstructing}. The merger history of the MW, following this early ancient massive merger, is expected to be far more quiet \citep[e.g.][]{deason2013broken,naidu2020evidence,Evans2020}.

The discovery of many populations of distinct halo-like stars, with high radial velocity anisotropy, has led to the conclusion that the GSE deposited debris onto the MW with a highly eccentric orbit. For example, \citet[][]{iorio2021chemo} presented a dominant sample of RR Lyrae with orbital anisotropy $\beta \sim 0.9$, within $5 < R \: [{\rm kpc}] < 25$. The inner portion of the stellar halo is also found to contain a relatively metal-rich population of blue horizontal branch (BHB) stars with high radial anisotropy which point to the GSE contributing 50\% of the Galactic halo \citep[][]{lancaster2019halos}. Numerous other studies support these claims of a collection of high eccentricity halo stars which owe their origin to a past merger \citep[e.g.][]{necib2019inferred, bird2021constraints}.

{\it N}-body simulations of high mass ratio $(q \gtrsim 0.1)$, low initial circularity $(\eta \lesssim 0.5)$ mergers show that satellites with such properties are prone to sinking deep within their host, and to radialization -- the process of dramatically increasing the satellites' eccentricity via dynamical friction, self-friction and the displacement of the centre of density relative to the centre of mass as the host moves towards the satellite's orbital pericentre \citep[][]{amorisco2017,naidu2021reconstructing,vasiliev2022radialization}. During these events, large satellites deposit their debris in distinct episodes of stripping. Each ``stripping episode'' has a unique mean energy, unique angular momentum, and unique energy spread. These properties are set at the point when the progenitor reaches pericentre due to the sinking and radialization. In a dramatic GSE-like event, the debris is deposited fast, in large quantities, and therefore the number of stripping episodes is small ($\sim 3$ -- $4$). Another feature attributed to such a high mass, high eccentricity, and rapidly dissolving merger is the ``pile-up'' of apocentric radii by stars deposited together, likely in the dominant stripping episode \citep[][]{deason2013broken, deason2018apocenter}. Since stars slow down at their apocentre, a collection of stripped stars on similarly eccentric orbits will bunch up together and form shells around their host \citep[][]{Quinn1984, hendel2015tidal}. This apocentre pile-up has been linked to the discovered break radius in the galactic halo, at around $r \sim 20$ kpc, where the density transitions from a shallower power law profile of $\rho(r) \sim r^{-2.5}$ to a steeper one $\rho(r)\sim r^{-4}$ \citep[][]{watkins2009substructure, sesar2011shape, xue2015radial, deason2018apocenter, iorio2019shape}. More recent work suggests that the stellar halo density may be fit better by a doubly-broken power law, with breaks at about $15$ and $30$ kpc \citep[][]{naidu2021reconstructing,han2022stellar}. While the GSE dominates the inner stellar halo, simulations show that its stars should extend over a wide range of Galactocentric radii. 

The merger's debris cloud, while nebulous and shapeless in configuration space, appears finely substructured in phase space due to the process of \textit{phase mixing}, similar to the formation of a stellar stream. While mostly unchanged in integral of motion space, it is now well understood that once debris is stripped from its progenitor onto the host, its distribution in phase space constantly evolves over time \citep[see][Section 4.10.2]{binneyandtremaine2008}. When bound to the progenitor, the satellite's stellar material is compact in phase space. However, when tidally stripped onto the host, small differences in orbital frequencies from the initial velocity dispersion cause the phase space distribution to stretch and wind up as it phase-mixes. As the satellite debris makes multiple passages around its host, it eventually folds in on itself. 
For a highly radial merger this evolution is seen most clearly in radial velocity versus radius or $(v_r,r)$ space. The satellite debris expands in the radial direction as it is spread across the host, and eventually begins to wrap and wind up into a series of ever-thinning chevrons as the debris continues to orbit around the host \citep[e.g.][]{bertschinger1985self, fillmore1984self, sanderson2013shells}. Each stripping episode of a merger presents a separate set of chevrons, with unique average energy and average angular momentum. Recently, signatures reminiscent of these phase space chevrons have been found in the MW's stellar halo around the Sun using \textit{Gaia} DR3 data \citep[][]{belokurov2022energy}.


It has been shown that these radial phase space chevrons can be matched onto stripes in energy versus radial angle or $(E, \theta_r)$ space \citep[][]{dong20226dshells}. The behaviour and appearance of the phase mixed debris is far simpler in this space, as the energies of the stars are approximately conserved. After accretion, the debris belonging to a single stripping episode is initially spread out in energy, but compact in radial angle. However, the distribution immediately starts to widen in the radial angle space, with the initially spread energy distribution imprinted. Since the frequency in the radial direction ($\dot{\theta}_r \equiv \Omega_r$) increases for decreasing energy, the distribution in $(E, \theta_r)$ space eventually becomes sheared and form stripes which are tilted by an increasingly smaller angle from the horizontal. Each stripe corresponds to a chevron in $(v_r, r)$ space. While the chevrons are intricately wound-up in phase space, their appearance in the $(E,\theta_r)$ space is much simpler, and this motivates us to introduce a novel method for quantifying disturbances by performing a 2-d Fourier transform and considering the power spectrum in this space. From this power spectrum, we obtain the \textit{ironing} parameter, which is discussed in detail in Section \ref{quantifying}. This new parameter is primarily aimed at quantifying the impact of perturbers on simulated chevrons and is not currently intended for use in observational data, given the uncertainties in stellar phase-space coordinates and in the Galactic potential.

As we will show, these $(v_r,r)$ chevrons, and hence also the $(E, \theta_r)$ stripes, respond to interactions of DM subhaloes by becoming blurred, smoothed, or entirely destroyed. The cold and fine grained nature of the phase mixed debris makes it sensitive to perturbation by Galactic subhaloes. The large radial extent of the debris means that, in principle, the phase space substructure of a major merger can act as a vast net to catch DM subhaloes with, behaving like a detector with a far larger cross-section than any individual stellar stream. This may allow us to constrain the nature of not only individual subhaloes, but also populations of subhaloes which span the extent of the merged debris. Since the GSE event occurred early in the history of the MW, one may expect the phase space substructure might be already blurred and undetectable. As per the CDM model, the formation of a typical galaxy is a long and turbulent process plagued by constant \& numerous perturbations \citep[][]{white1991galaxy}. However, the assembly history of the MW, dominated by a single ancient massive merger with a much more gentle subsequent merger history, makes it fortunately suited to the kind of detection methods we discuss.

While no other galaxy has had the influence on our own Galaxy's composition like the GSE has, the continued formation of the MW via mergers is evident by the large $(N>50)$ collection of known dwarf galaxies around it \citep[see, e.g.][for orbital properties]{pace2022proper}. However, most of these dwarf galaxies' orbital properties mean they are unlikely to visibly disturb the GSE debris, since the dwarfs with the most massive (known) masses of $M\sim 10^7$--$10^8$ M$_{\odot}$ have pericentres greater than 40--50 kpc \citep[e.g.][]{simon2007kinematics, lokas2009mass, pace2022proper}. The two key exceptions to this are the uniquely massive Large Magellanic Cloud (LMC) and the Sagittarius dwarf spheroidal galaxy (Sgr). The LMC has a total mass of around $(1-2) \times 10^{11}$ M$_{\odot}$, likely just past its first approach pericentre of about 50 kpc about the MW \citep[][]{kallivayalil2006smc, erkal2019total}. Given the LMC's close proximity as the MW's largest satellite, its effect on the Galactic halo has already been shown to be substantial \citep[][]{erkal2021detection}. While a few orders of magnitude less massive at present, Sgr's smaller pericentre makes it an arguably more impactful perturber of the GSE debris. Originally discovered by \citet[][]{ibata1995sgr}, Sgr has also just passed its most recent pericentre of around 16 kpc, and with its present-day mass of $\lesssim 5\times10^8\,M_\odot$ is expected to dissolve over the next few billions years \citep[][]{vasiliev2020last}. However, its mass at earlier times was surely much higher, around $10^9$--$10^{10}$ M$_{\odot}$ \citep[][]{niederste-ostholt2010re,bennett2022exploring}, or even up to $\gtrsim 6\times10^{10}$ M$_{\odot}$ \citep[][]{jiang2000sgr,gibbons2017sgr}. The unique behaviour of the LMC and Sgr means they may be able to have their orbits or total masses constrained by the discovered local phase space chevrons.

The outline of this work is as follows. In Section~\ref{background}, we briefly explain the mechanism of formation of substructures in the $(v_r, r)$ and $(E, \theta_r)$ spaces with the aid of an isolated idealised merger. We then follow this up with details of our method for quantifying the subhalo impact on simulated chevrons in $(v_r,r)$ space in Section~\ref{quantifying}. Next we detail a more sophisticated {\it N}-body simulation and test-particle integration in Section~\ref{simulations}, which are used to generate our results. In Section~\ref{genericsubhaloes} we explore the results of these simulations involving perturbing subhaloes and their impact on the GSE debris substructure. Lastly in Section~\ref{conclusions} we summarise our results and consider future work.

\section{Background}\label{background}

\begin{figure}
    \centering
    \includegraphics[width=0.48\textwidth]{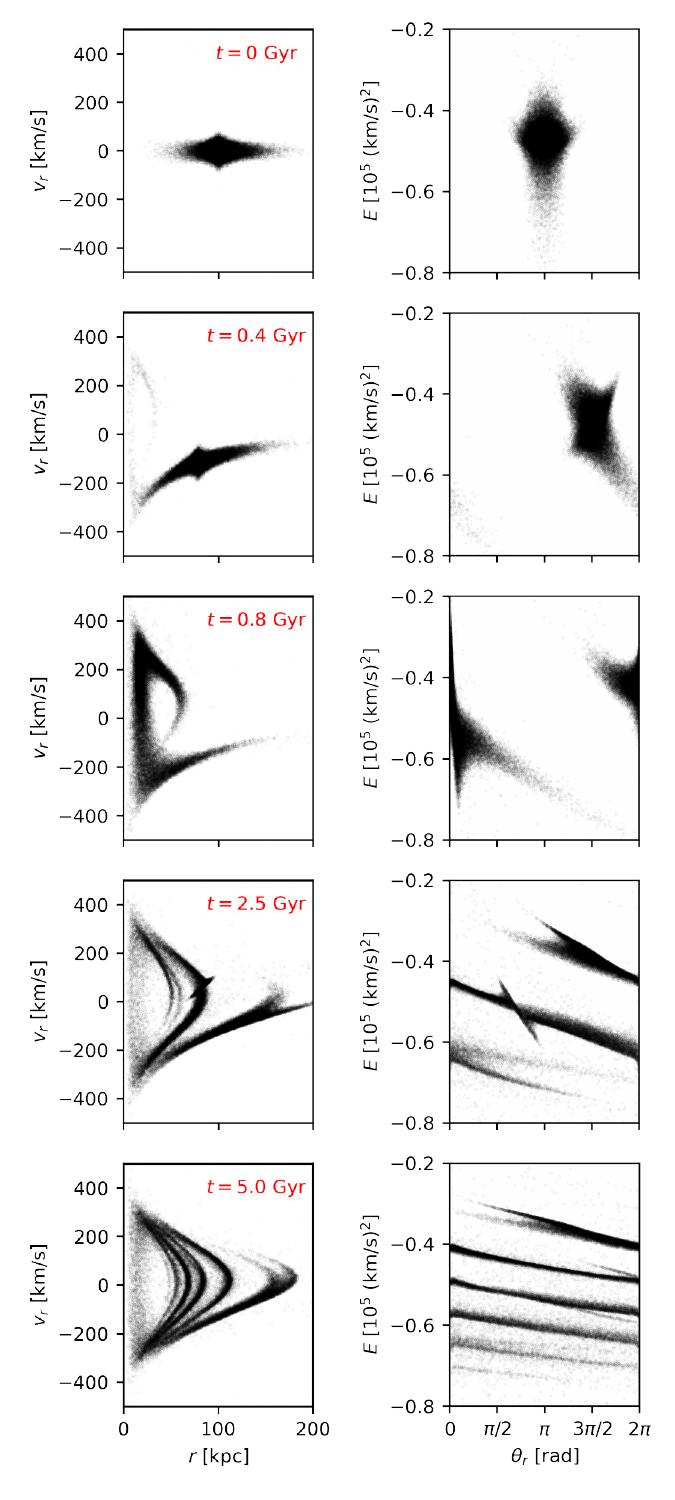}
    \caption{Snapshots of an idealised merger, described in Section~\ref{background}, in which a self-gravitating progenitor is accreted onto a static host potential. We show a scatter plot of the $(v_r,r)$ chevrons in the left column and the corresponding $(E,\theta_r)$ stripes in the right column, where each left-right pair presents a single time step. There are the same number of chevrons in the left as number of stripes in the corresponding right panel.}
    \label{fig:toy_model}
\end{figure}

Here we describe the mechanism behind the formation of the $(v_r,r)$ chevrons and the $(E,\theta_r)$ stripes. As a visual aid, we present snapshots from an idealised merger which lasts 5 Gyr. In Figure~\ref{fig:toy_model}, we show the chevrons in the left column and the stripes in the right column at various illustrative time steps in order to show the formation of the substructure as described in this section. The idealised model consists of a self-gravitating progenitor that is a truncated NFW with mass $M=10^{10}$ M$_{\odot}$ and concentration $c=10$, represented by $5\times10^4$ particles. The host galaxy is represented by a fixed potential profile (\textsc{MilkyWayPotential} from \textsc{Gala}, \citealt{price-whelan2017gala}). The idealised merger presented in this section is not referenced beyond this section and is only for pedagogical use.

\subsection{Phase mixing in \texorpdfstring{$(v_r,r)$}{(vr,r)} coordinates}

\begin{figure}
    \centering
    \includegraphics[width=0.48\textwidth]{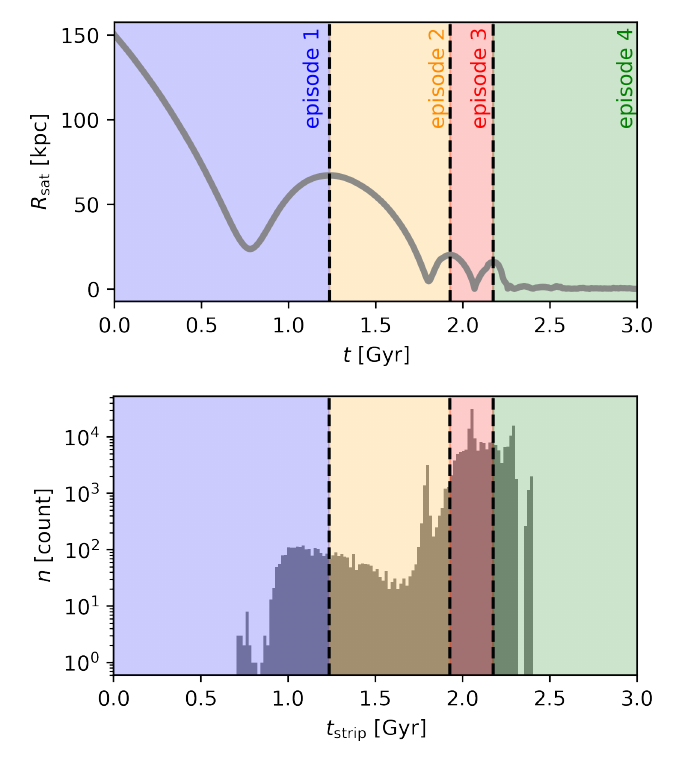}
    \caption{Stripping episodes and stripping times of the satellite star debris from the more realistic $N$-body simulation described in Section~\ref{simingthesausage}. {\it Top:} Trajectory of the satellite in the frame of the centre of the host for the first 3 Gyr of the {\it N}-body simulation. The trajectory is split up by a series of stripping episodes, defined as the times between each apocentre of the satellite. {\it Bottom:} Histogram showing the number of satellite stars stripped during each stripping episode. A star is considered ``stripped'' from the satellite progenitor when its energy relative to the progenitor first exceeds zero.}
    \label{fig:stripping_times}
\end{figure}

\begin{figure*}
    \centering
    \includegraphics[width=0.95\textwidth]{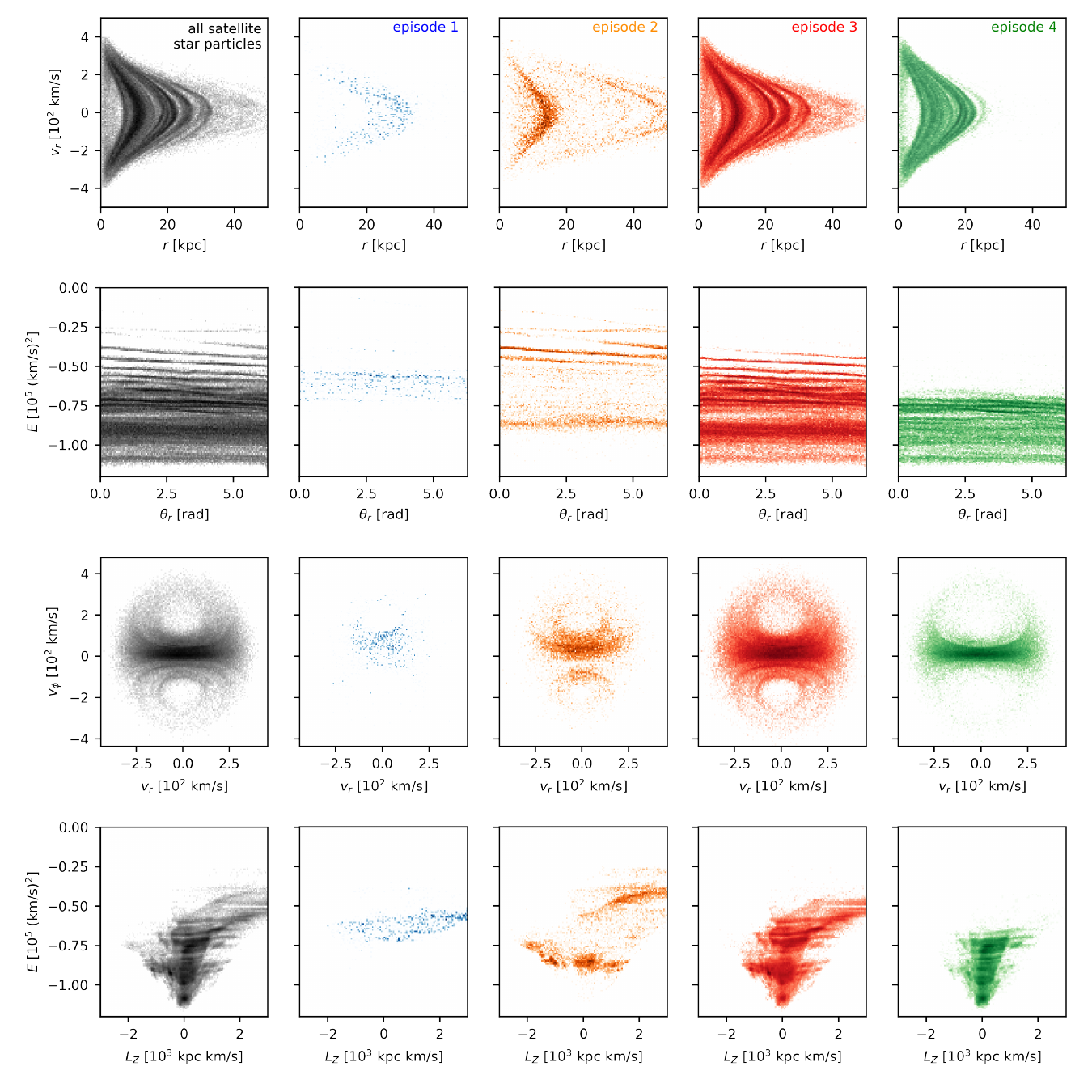}
    \caption{Final snapshot of the satellite debris broken down by stripping episodes, from the more realistic $N$-body simulation described in Section~\ref{simingthesausage}. {\it Top row:} $(v_r, r)$ coordinates for the debris in totality (left column, in grey) and for each stripping episode (in colour). Note that the final two episodes dominate, as additionally illustrated in Figure~\ref{fig:stripping_times}. {\it Top, middle row:} $(E, \theta_r)$ coordinates for the debris. Each stripping episode covers a different range of energy and the stripes appear at varying frequency. {\it Bottom, middle row:} $(v_{\phi}, v_r)$ space of the debris, illustrating the likeness of the merger to the GSE merger. Episode 3 clearly dominates in the appearance of an ellipsoid feature. {\it Bottom row:} $(E, L_Z)$ space of the debris, where the separation between leading and trailing trails of the merger is most visible in episodes 2 and 3.}
    \label{fig:episode_dynamics}
\end{figure*}

In a sufficiently high mass-ratio and high-eccentricity merger event, debris will be tidally stripped from an infalling satellite onto its host and form shell-like structures in configuration-space \citep{Quinn1984, hendel2015tidal}. Before the satellite merges with its host, its stars are confined to a compact region of phase space (first row of Fig.~\ref{fig:toy_model}). However, as the satellite debris is stripped, its phase space distribution evolves. The initially compact region of phase space now covers a wider spatial extent as it spreads out across the host, causing the velocity distribution to thin out according to Liouville's theorem (second row of Fig.~\ref{fig:toy_model}). Subsequently, small differences in the orbital frequencies of stars cause further evolution of the phase space via phase-mixing; the debris material continues to stretch and thin out, and eventually the phase space distribution folds in on itself and winds up in a spiral as stars complete numerous orbits. For a highly eccentric GSE-like merger, this phase-mixing manifests most clearly in Galactocentric spherical polar $(v_r, r)$ space as a series of chevrons (fifth row of Fig.~\ref{fig:toy_model}). It has been shown by \citet[][]{sanderson2013shells} that the chevrons can be approximately fit by the quadratic function near their apocentre
\begin{equation}\label{eq:chev_quadratic}
    r = r_s - \kappa (v_r - v_s)^2,\qquad\qquad
    \kappa \simeq  (2\,\mathrm{d}\Phi/\mathrm{d}r)^{-1},
\end{equation}
where $r_s$ is the chevron's maximum radius, $v_s$ is the radial velocity at $r_s$, and $\kappa$ is related to the gradient of the host potential.

The large core of the GSE progenitor can survive numerous pericentre passages, and therefore deposit stars in a several distinct ``stripping episodes'', which often happens in non-uniform time intervals \citep{dupraz1987dynamical}. For example, note the continued presence of a compact progenitor clump in $(v_r,r)$ space at about $r=100$ kpc in the fourth row of Fig.~\ref{fig:toy_model}. Each star possesses a unique ``stripping time'' $t_{\rm strip}$, which we define as the time at which the star's energy relative to the satellite potential first becomes positive. The time-spans of each stripping episode is set by the times between apocenters of the infalling satellite progenitor. The time-span of each episode of our initial GSE-like merger, and the associated number of satellite stars, can be seen in Figure~\ref{fig:stripping_times}. Since shells can possess different velocities at the same radii, it is almost certain that the phase space chevrons for different stripping episodes will overlap, merge, and form what have been dubbed ``super-chevrons''. 

As the phase-mixing continues, the chevrons become thinner and more frequent along the radial direction eventually causing the phase space structure to appear uniform. Therefore, if a simulation is too low resolution, this may hide the true nature of the phase space folding. The complicated nature of the overlapping chevrons in each episode is somewhat simplified by binning the stars into stripping episodes (see Figure~\ref{fig:episode_dynamics}). 

Provided there is high enough resolution, the phase space folds can be used to deduce the age the merger event; the longer the debris phase-mixes, the more phase space chevrons will form.

\subsection{Phase-mixing in \texorpdfstring{$(E, \theta_r)$}{(E,thetar)} coordinates}

The structure of the shells can be also explored by considering action-angle coordinates, specifically in $(E,\theta_r)$ space, as presented in the right column of Figure~\ref{fig:toy_model}. A thorough explanation of action-angle coordinates can be found in chapter 3.5 of \citet[][]{binneyandtremaine2008}. Given any time-independent, integrable potential $\Phi$ for the Galactic host, and thus a  Hamiltonian $H$, we can find canonical coordinates consisting of actions $\bm{J}$ and angles $\bm{\theta}$. Crucially, the actions $\bm{J}$ are integrals of motion. Such coordinates are defined so that 
\begin{equation}\label{eq:actangdefinition}
    \dot{\bm{\theta}} = \frac{\partial H (\bm{J})}{\partial \bm{J}} \equiv \bm{\Omega}(\bm{J}), \:\:\:\:
    \dot{\bm{J}} = -\frac{\partial H (\bm{J})}{\partial \bm{\theta}} = 0,
\end{equation}

where ${\bm \Omega}({\bm J})$ are the constant frequencies of the angles, i.e. $\theta_i(t)~=~\theta_i(0)~+~\Omega_i t$. The dynamics of action-angle coordinates are therefore reasonably simple; the actions are constant and the angles increase linearly in time. The actions have the property of adiabatic invariance. They are approximately preserved under slow changes to the potential, such as gradual accretion of mass. In 6-d action-angle space, a bound orbit moves in the three $\bm{\theta}$ directions, over a 3-torus defined by the three actions $\bm{J}$. In a spherical potential, the Hamiltonian admits three independent actions in the radial, azimuthal, and vertical direction for each orbit:
\begin{align}
    J_r &= \frac{1}{\pi}\int^{r_{\rm apo}}_{r_{\rm peri}} \big[2\big(E-\Phi(r)\big) - L^2/r^2\big]^{1/2}\,\mathrm{d}r \\
    J_{\phi} &\equiv L_z \\
    J_z &\equiv L - |L_z|,
\end{align}
and in realistic axisymmetric potentials most orbits conserve three actions with the same physical meaning (though the expressions for $J_r$ and $J_z$ are more complicated).
Each action describes the extent of oscillation of an orbit in each direction. The three corresponding angles are ${\bm \theta} = (\theta_r, \theta_{\phi}, \theta_{\theta})$, which are defined by equation \ref{eq:actangdefinition}. Note that $\theta_r = 0$ is defined as pericentre and $\theta_r = \pi$ is apocentre. For a highly eccentric orbit ($L\sim 0$), the frequency in the radial direction is approximately a function of energy alone,
\begin{equation}\label{eq:omega}
    \Omega_r(E) = \frac{2\pi}{T_r(E)},
\end{equation}
where the radial period is
\begin{equation}
    T_r(E)~=~2\int^{r_{\rm apo}}_0 \big[2(E-\Phi)\big]^{-1/2} \mathrm{d}r.
\end{equation}

The behaviour of the shells in $(E, \theta_r)$ space is more straightforward than in $(v_r, r)$ space, as introduced and explained in detail in \citet[][]{dong20226dshells}, wherein more illustrative visualisations are found. While the shells take the form of numerous wrapping chevrons in $(v_r, r)$ space, in $(E, \theta_r)$ space they flatten out to form stripes (at a slight angle to the horizontal), as shown in the right column of Figure~\ref{fig:toy_model}. Each chevron in phase space can be matched to a stripe in action-space. Particles on approximately energy conserving orbits appear as moving only horizontally in $(E, \theta_r)$ space at constant speed $\Omega_r(E)$, given by equation \ref{eq:omega}. 

The formation of the stripes is as follows. If the satellite moves at a speed $v$, and the velocity dispersion of particles bound to the satellite is $\sigma_v$, then the kinetic energy dispersion of these bound particles is of the order $\sim (v + \sigma_v)^2/2 - v^2/2 \simeq v\sigma_v$. Therefore, when the progenitor is at its maximum orbital speed at pericentre ($\theta_r = 0$), just prior to stripping, the energy dispersion of the satellite particles is at its largest. This results in a large vertical range in $E$ (see third row of Fig.~\ref{fig:toy_model}). This enlarged energy spread is then imprinted on the stripped particles. Following from this, since $\Omega_r (E)$ increases monotonically for lower energies, particles with more tightly bound orbits will move faster in $\theta_r$, resulting in a horizontal shearing. This causes the initially thin $\theta_r$ distribution to widen, and as the stars complete multiple orbits the stripe wraps around in $\theta_r$ (see fifth row of Fig.~\ref{fig:toy_model}). 

As with the phase space chevrons, this wrapping theoretically allows us to date the merger by counting the number of wrappings in $\theta_r$ \citep[][]{mcmillan2008disassembling,gomez2010identification}. Another feature in common with the chevrons is the increasing complexity of the distribution when the progenitor survives multiple passages and deposits debris in multiple episodes. The resulting picture is multiple overlapping, wrapping stripes. Figure~\ref{fig:episode_dynamics} illustrates this for the {\it N}-body simulation by breaking down the population of stellar debris by episode.

Throughout this work conversion to action-angle coordinates from cartesian coordinates is done using \textsc{Agama} \citep{vasiliev2019agama}, which uses the St{\"a}ckel fudge method \citep[][]{binney2012actions} to compute the actions in axisymmetric potentials.

\section{Quantifying the Subhalo Impact}\label{quantifying}

\begin{figure*}
    \centering
    \includegraphics[width=0.95\textwidth]{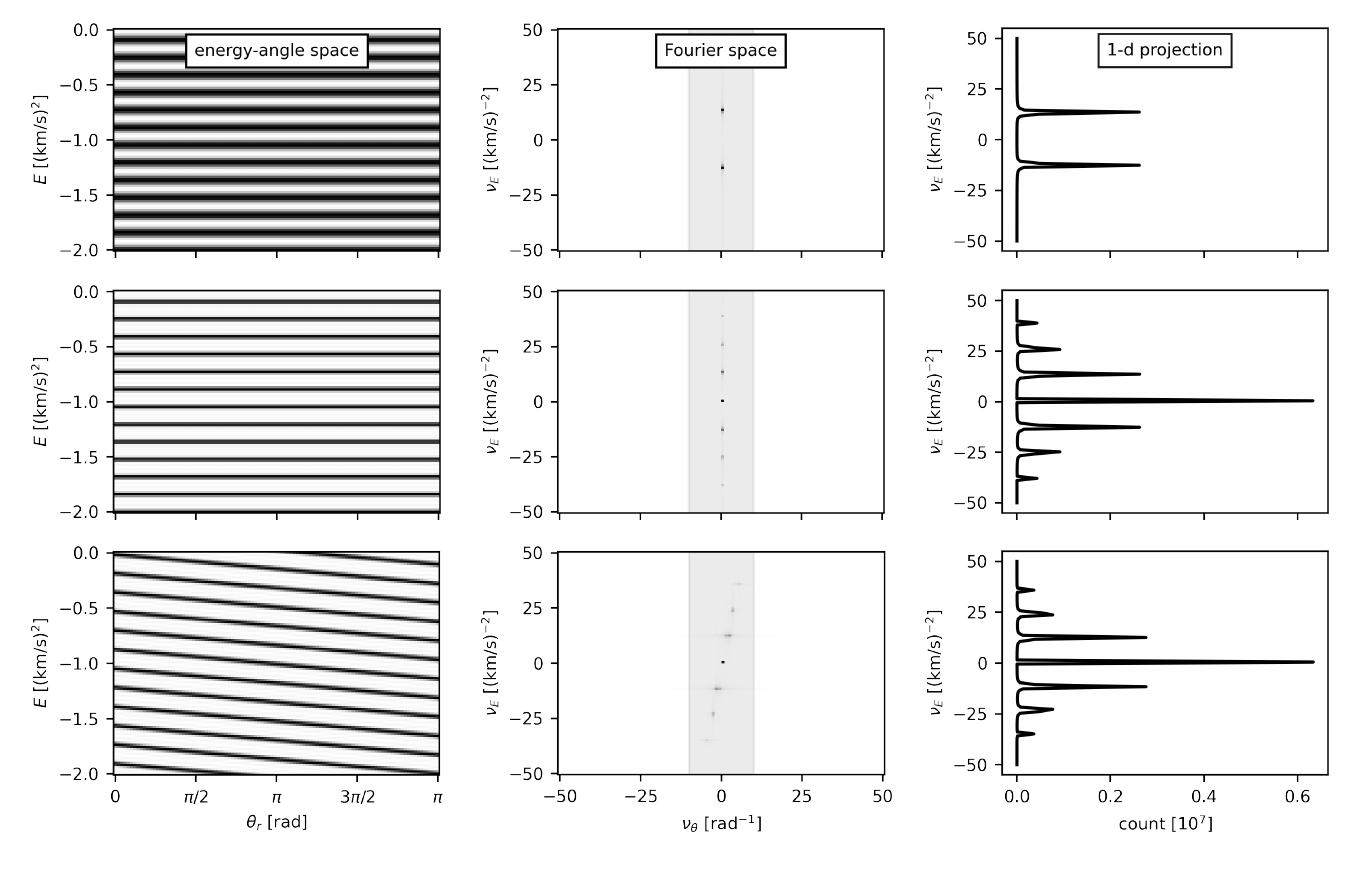}
    \caption{The 2-d Fourier transform of a series of artificial stripes, representative of the $(E,\theta_r)$ distribution after significant phase-mixing. For three examples, we show the stripes in the left column, the 2-d Fourier transform in the middle column, and a 1-d projection in the right column. {\it Top:} The stripes are constructed by 2-d cosine wave with one fixed frequency. In Fourier space, these stripes appear as points (one positive, one negative) at a given energy frequency. These points appear as peaks in the 1-d projection. {\it Middle:} Here we construct the stripes using a superposition of three 2-d cosine waves, each with different frequencies. This translates to three identifying frequencies in both Fourier space and the 1-d projection. Note that the points fall along a vertical line in Fourier space. {\it Bottom:} Lastly we take the same three 2-d cosine waves, but tilt them at an angle, which is more representative of the genuine $(E, \theta_r)$ stripes. In Fourier space, this now translates to a series of points, arranged along a line, at an angle to the vertical.}
    \label{fig:mock_ft}
\end{figure*}

\begin{figure}
    \centering
    \includegraphics[width=0.48\textwidth]{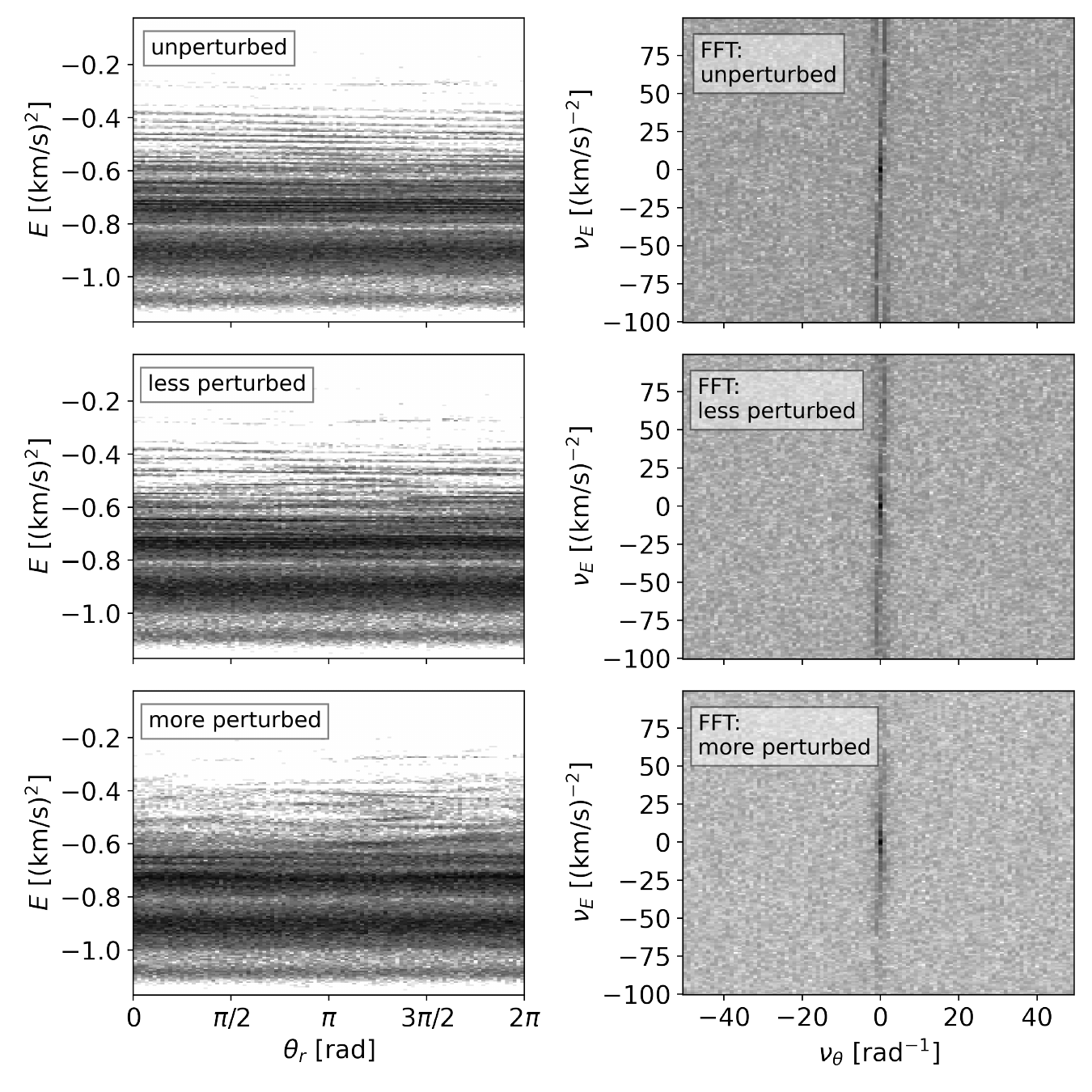}
    \caption{2-d Fourier transform of $(E,\theta_r)$ space to energy-angle frequency $(\nu_E, \nu_{\theta})$ space, for the example cases of a $10^{9}$ (less perturbed) and $10^{10}$ M$_{\odot}$ (more perturbed) subhalo that disturb the phase-mixed satellite debris for 5 Gyr. {\it Left:} The energy-angle space space, showing clear damage done to the high energy $(E \gtrsim -0.6)$ stripes for the two perturbed cases. {\it Right:} The corresponding power spectra. Note the change of structure in the primary feature in the power spectra from the unperturbed to the perturbed case.}
    \label{fig:fm1}
\end{figure}

\begin{figure}
    \centering
    \includegraphics[width=0.48\textwidth]{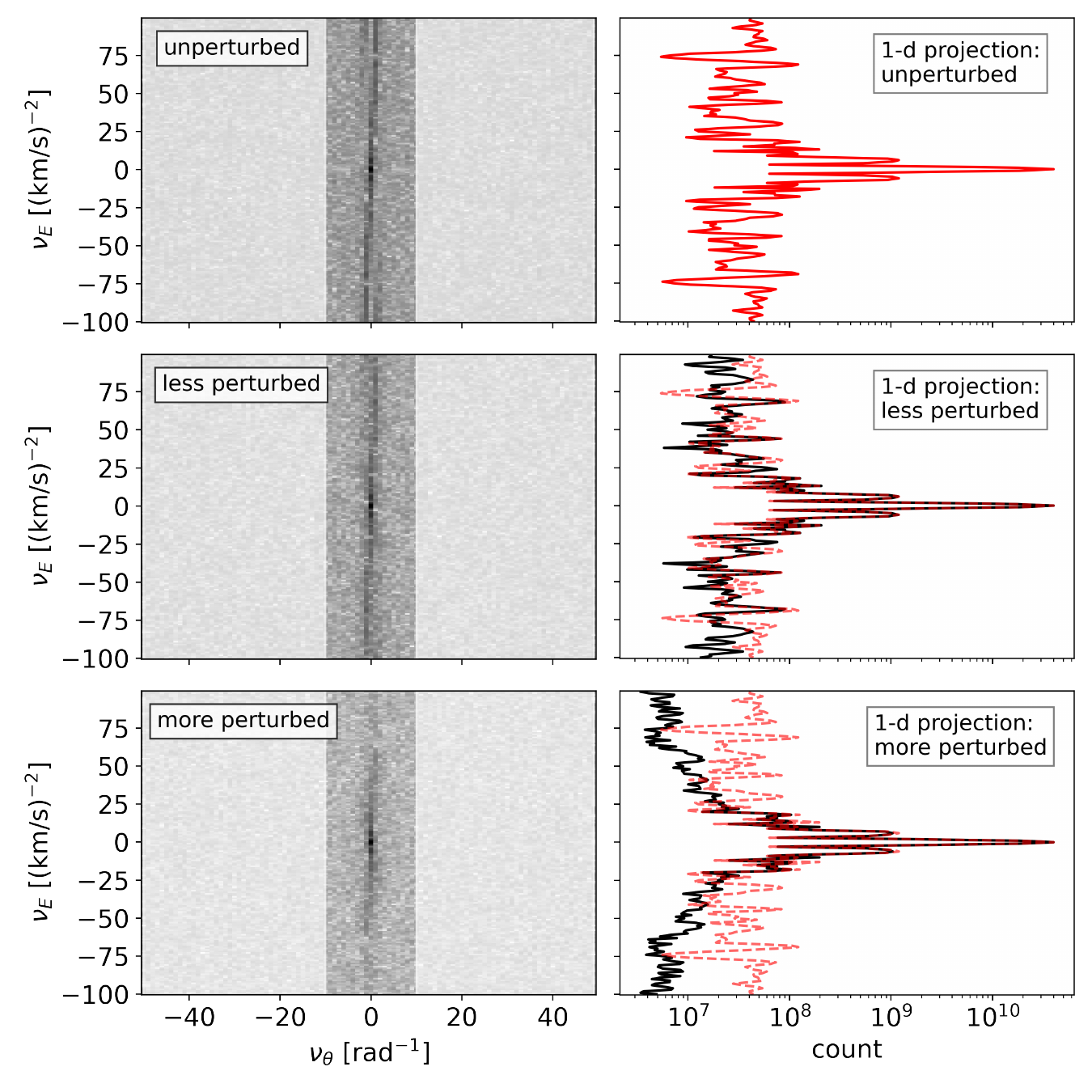}
    \caption{1-d projection of the 2-d power spectra to illustrate the change in energy frequencies for the example cases of two perturbers: a $10^{9}$ (less perturbed) and $10^{10}$ M$_{\odot}$ (more perturbed) subhalo. {\it Left:} The ironing parameter is only calculated for specific range of low-frequencies angles to minimise noise. The excluded range of frequencies are those faded out. {\it Right:} The 1-d projection of the power spectra. The unperturbed signal is shown in red in all three panels, while the relevant perturbed signal is shown in black. Note the loss of power in the high energy frequencies in the perturbed cases, and the more dramatic impact in the ``more perturbed'' scenario.}
    \label{fig:fm2}
\end{figure}

\begin{figure}
    \centering
    \includegraphics[width=0.48\textwidth]{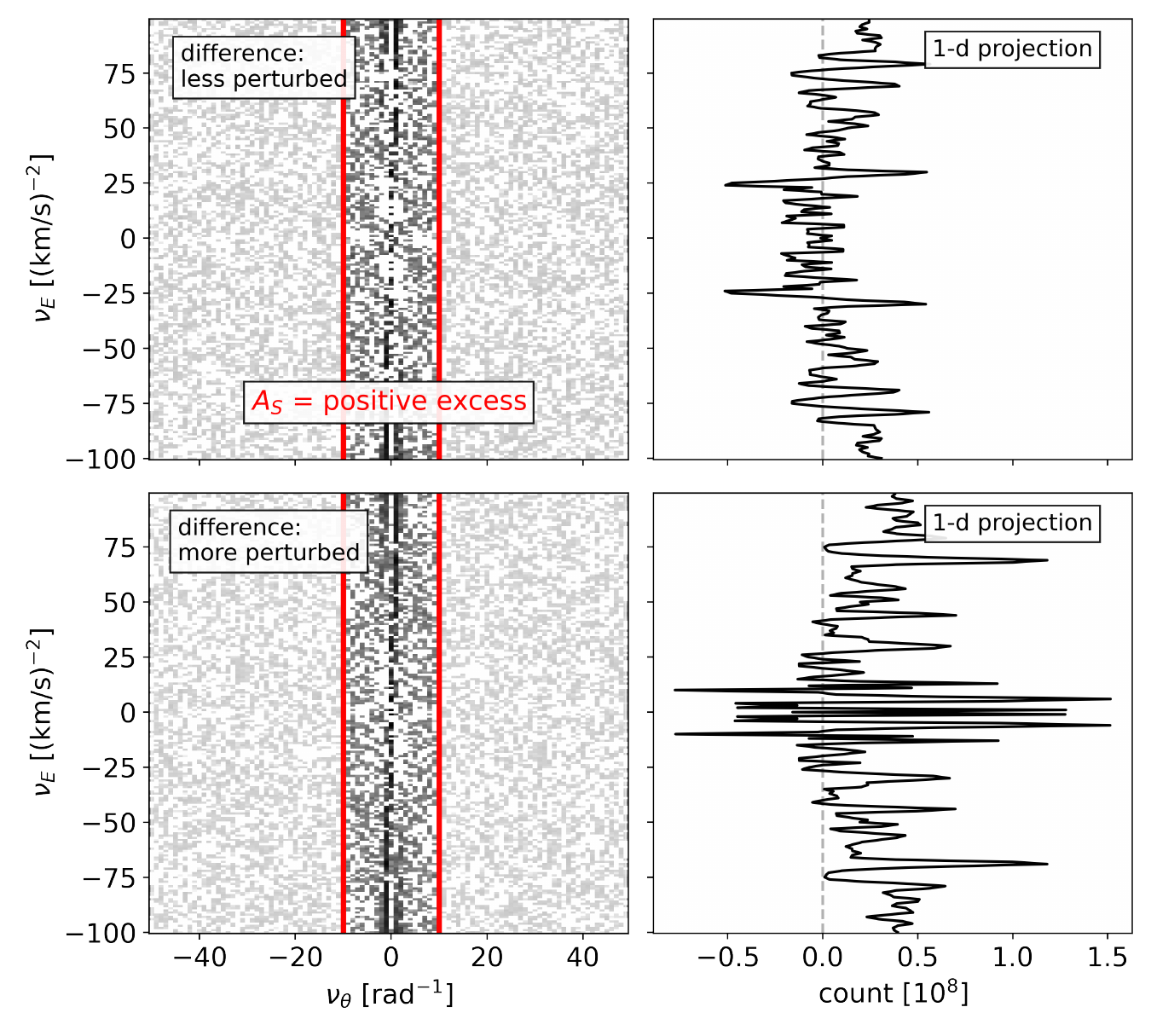}
    \caption{The difference between the 2-d power spectra ({\it left}) and the 1-d projection of this difference ({\it right}) for the example cases of two perturbers: a $10^{9}$ (less perturbed) and $10^{10}$ M$_{\odot}$ (more perturbed) subhalo. We calculate the ironing parameter $A_S$ by summing the positive excess of the difference between the power spectra, in the low theta frequencies (the region between the red lines).}
    \label{fig:fm3}
\end{figure}

In this section, we describe the calculation of the ironing parameter, $A_S$, which is used as a quantitative measure of the impact of a perturbing subhalo on our simulated phase-mixed satellite debris. Firstly, it is essential to transform from $(v_r,r)$ space to energy-angle space to obtain a more clean representation of the phase mixed debris. At a given time step, we then compare the $(E,\theta_r)$ \textit{frequency} distribution of the satellite debris as it evolved in the static host potential \textit{without} perturbation, with the distribution as it evolved in a time-dependent potential \textit{with} a perturbing subhalo. The method described in this section is graphically presented in Figures~\ref{fig:fm1} -- \ref{fig:fm3}. With the aid of these Figures, we detail the calculation of the ironing parameter for a model example of a subhalo. To calculate $A_S$ at a given timestep, there are four basic steps:

\begin{enumerate}
    \item 2-d Fourier transform the $(E,\theta_r)$ space into energy and angle frequencies $(\nu_E, \nu_{\theta})$, for {\it both} the unperturbed debris and the perturbed debris, in order to produce 2-d power spectra of the frequencies;
    \item reduce noise by masking out the high $\theta_r$ frequencies of the 2-d power spectrum;
    \item subtract the perturbed power spectrum from the unperturbed one;
    \item sum the positive excess of the difference between the perturbed and unperturbed power spectra to measure the decrease in power.
\end{enumerate}

The motivation for this method is as follows. Since the $(E,\theta_r)$ space is made up of a series of stripes, whose frequencies along the energy axis relate to the time elapsed since the merger, it seems reasonably intuitive to 2-d Fourier transform this space into the space of energy frequencies $\nu_E$ and angle frequencies $\nu_{\theta}$. For example, Fig.~\ref{fig:mock_ft} presents a series of mock distributions which are representative of the $(E,\theta_r)$ space of actual merger debris, alongside their 2-d Fourier transform. In this Figure, we show how a series of horizontal stripes of varying frequency corresponds to points in 2-d Fourier space which fall along a central vertical line. However, when the original stripes are set at an angle to the horizontal, the Fourier space points now appear along a line which is at that same angle, but to the vertical. By integrating the 2-d power spectrum over the low-frequency portion of angle frequencies $\nu_\theta$, we project onto a 1-d power spectra. Here we see a set of peaks which corresponding to the frequencies of the stripes in the original distribution. These 1-d power spectra are shown for illustrative purposes only, and are not used in calculations. Any process that results in a change of frequency of the stripes will correspond to a shift in power in the 2-d power spectra (and will also be seen in the 1-d power spectra). In Figures~\ref{fig:fm1} -- \ref{fig:fm3}, we present a genuine example of the $(E, \theta_r)$ space resulting from the phase mixing of the {\it N}-body simulation and how it presents in 2-d Fourier space. It is worth noting that, in reality, the tilt angle becomes smaller for higher binding energies. Therefore, the stripes will not be exactly parallel. All 2-d Fourier transforms are computed using the \textsc{numpy.fft} package, from the 2-d $(E,\theta_r)$ histogram with 200 $y$-bins and 100 $x$-bins. The same bins are used throughout the entire work for all plots of $(v_r,r)$ and $(E, \theta_r)$.

Step (i) is illustrated in Figure~\ref{fig:fm1}. Upon examination of the unperturbed power spectrum (top row), we can see a dominating feature alluded to by the mock example: a long (almost vertical) streak of varying density, where the density peaks along the line are indicative of the most prominent energy frequencies. The panels beneath present a ``less perturbed'' and ``more perturbed'' snapshot. The less perturbed case consists of a $10^{9}$ M$_{\odot}$ perturbing subhalo, whereas the more perturbed case is a $10^{10}$ M$_{\odot}$ subhalo on the same trajectory. Note the way in which the more perturbing case causes a very visible reduction in the higher energy frequencies $|\nu_E| > 50$ (km/s)$^{-2}$, while in the less perturbed case this reduction is barely detectable. The subhalo interaction causes a ironing-out of the energy stripes, resulting in power being dissipated from higher frequencies to lower frequencies. It is important to note that, because the frequency of the $(E, \theta_r)$ stripes is dependent on the time since the merger, power will be dissipated even over the course of an unperturbed simulation, but from a lower frequencies to higher frequencies. 

Step (ii) is presented in Figure~\ref{fig:fm2}. We isolate the dominating feature to reduce the pick-up of unnecessary noise in the higher angle frequencies $|\nu_{\theta}| > 10$ rad$^{-1}$. In the left column, we present the excluded regions as faded out. In the right column, we show the masked region of the power spectra projected onto the y-axis to make these density peaks clear. By comparison of the unperturbed power spectrum (red signal) with the perturbed power spectrum (black signal) in Figure~\ref{fig:fm2}, we can see a clear difference in the location of density peaks. Typically, after perturbation the higher frequency peaks become reduced in amplitude, or in an extremely violent case the peaks may be entirely removed. 

Step (iii) and (iv) are presented in Figure~\ref{fig:fm3}. To calculate $A_S$, we can subtract the 2-d perturbed power spectrum from the unperturbed power spectrum to see the impact that the perturbation has on the frequencies of the energy stripes. As our final measure of the subhalo impact, we sum up the positive excess of the difference between of the subtracted power spectra, within the region marked by the red lines. Again, the excluded angle frequencies are faded out. We sum the positive excess in order to simply calculate only how much power is lost compared to the unperturbed case. In our work, the excluded high angle frequency region is chosen as $|\nu_{\theta}| > 10 \: {\rm rad}^{-1}$. We denote the sum of the included region as $A_S$, which has rather unwieldy units of (km/s)$^{-2}$rad$^{-1}$. The right column of Figure~\ref{fig:fm3} shows the 1-d projection of the subtracted signals.

To beautify the numeric values of $A_S$, we divide by the raw difference by $10^{10}$, which is the order of magnitude of zero-frequency peaks in the power spectra. Therefore, all values of $A_S$ shown are actually $10^{10}$ times their presented value. Throughout this work we refer to $A_S$ as the \textit{ironing} parameter. We present illustrative values of $A_S$ in Section \ref{genericsubhaloes} to give an impression of what constitutes an impactful value of $A_S$.

These power spectra have a background of noise, which we need to estimate in order to understand the relative significance of $A_S$. To calculate noise $\sigma_A$ at a given time step $t_1$, we compare the unperturbed power spectrum at $t_1$ with the unperturbed power spectrum at $t_1-dt$, where $dt$ is the minimum time step in our simulations from $t_{\rm int}~>~(5+0)$ Gyr. In this case, we sum in quadrature and take the square root of the result to find $\sigma_A$. 

It is important to note that the ironing value alone is not enough to get a full picture of the impact of the subhalo on the $(v_r, r)$ chevrons or energy stripes. We include some illustrative figures to show the correspondence between $A_S$ and damage done to $(v_r,r)$. Should the absolute value of $A_S$ be comparable to $\sigma_A$, then we can be certain that the chevron substructure will remain almost entirely intact. In the case that $A_S \gg \sigma_A$, we can be certain that either some chevrons are destroyed, or the amplitude of the chevrons is reduced. However, for both intermediate and high ironing values, the $(v_r, r)$ space must be inspected to fully understand the damage done to substructure. We therefore treat the ironing value as a guiding measure of what order of magnitude impact to expect. Figure~\ref{fig:all_lineplot} and the appendix tables show that, all else being fixed, increasing the mass of a single perturbing subhalo by an order of magnitude corresponds to an increase in $A_S$ by roughly an order of magnitude.

\begin{figure}
    \centering
    \includegraphics[width=0.48\textwidth]{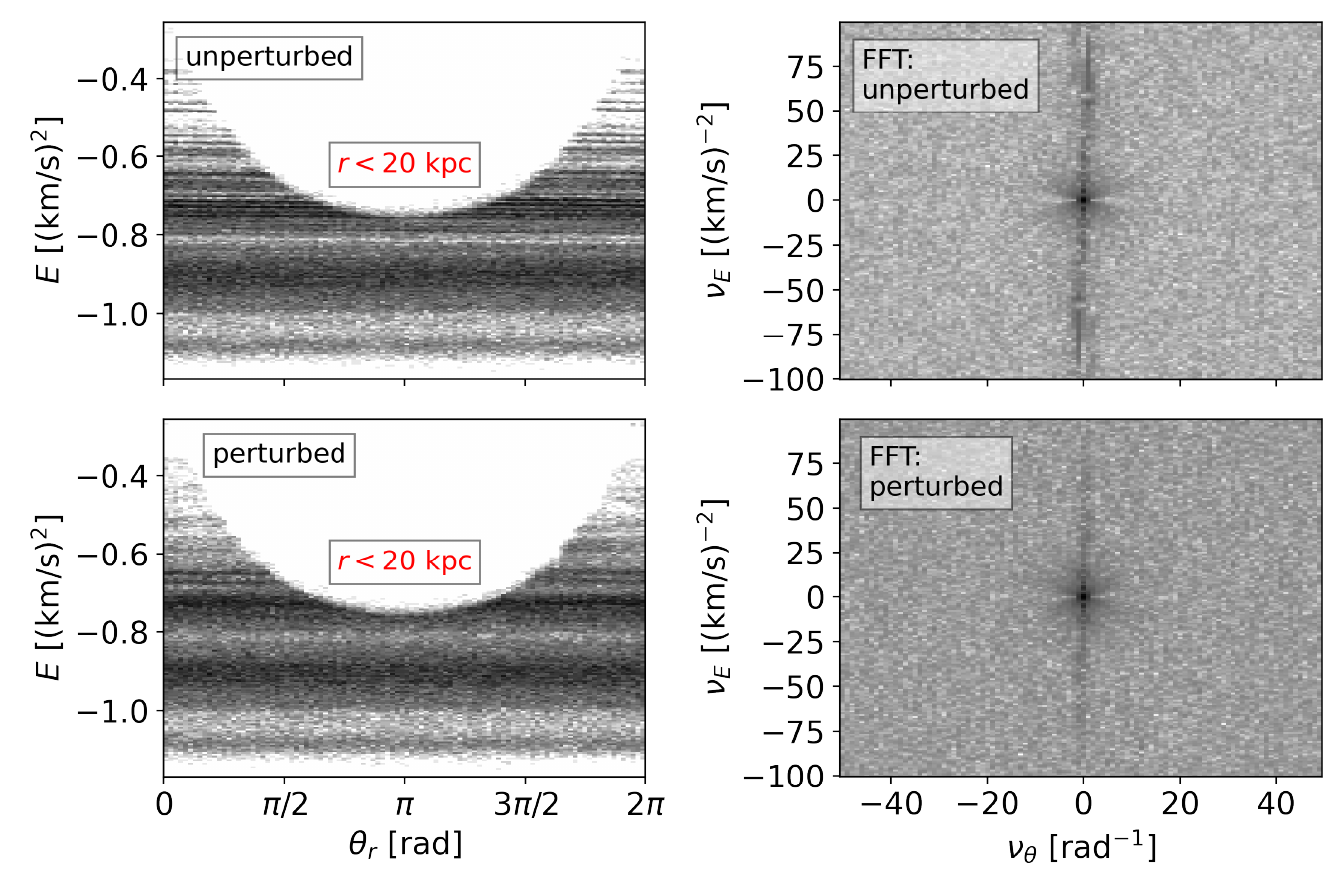}
    \caption{2-d Fourier transform of $(E, \theta_r)$ space to energy-angle frequency space, but with limited spatial extend $(r < 20 \: {\rm kpc})$. In this case, the perturbing subhalo has mass $10^{10}$ M$_{\odot}$ and pericentre $10$ kpc. By limit the spatial extent, we have excluded any high energy particles which are close to their apocentre. The widening of the Fourier space distribution, compared to the non-spatially limited case, is representative of the now clear variation in $\theta_r$ frequencies.}
    \label{fig:fm_r}
\end{figure}

\begin{figure}
    \centering
    \includegraphics[width=0.48\textwidth]{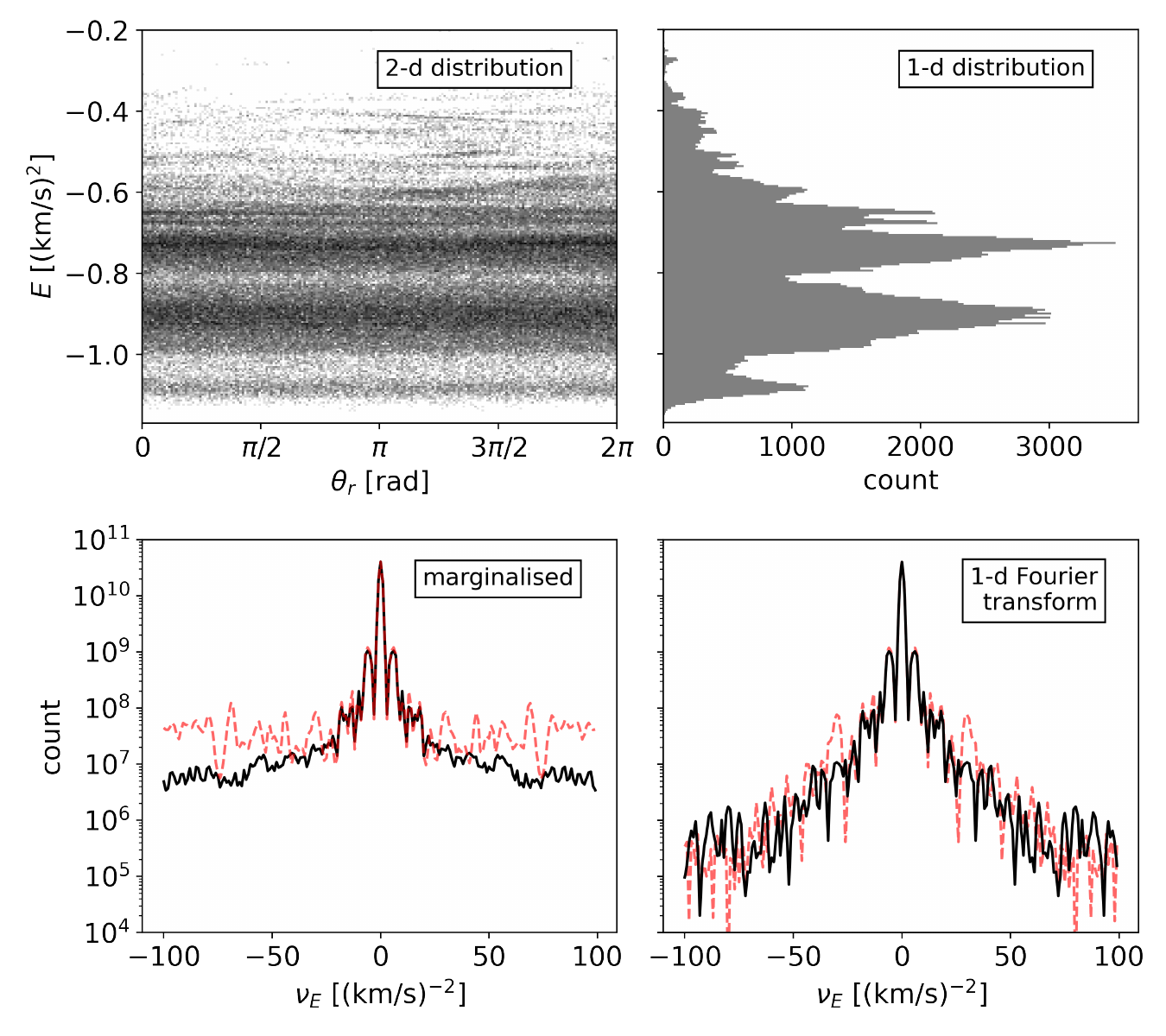}
    \caption{Comparison of the 1-d Fourier transform of the energy distribution with the 2-d Fourier transform and projected energy frequency peaks, for the same ``more perturbed" ($M=10^{10}$ M$_{\odot}$) case in previous plots. {\it Left:} The 2-d distribution of energies and angles (200 by 100 bins), and the associated marginalisation over low angle frequencies. {\it Right:} The 1-d distribution of energies with 200 bins, and associated Fourier transform. The red dotted line shows the unperturbed case, the black dotted line shows the perturbed case.}
    \label{fig:1d2d}
\end{figure}

Our method can also be applied to a spatially limited region of $(E, \theta_r)$ space. For example, in Figure~\ref{fig:fm_r} we present the energy-angle space and energy-angle frequency space for particles within $r<20$ kpc. There is an evident widening of the Fourier space distribution, compared to the non-spatially limited case. This is representative of the now clear variation in $\theta_r$ frequencies. Since a large portion of the high energy frequencies are removed, the sensitivity of $A_S$ to global changes in the chevrons is reduced. However, $A_S$ is still effective at measuring the ironing out of chevrons within 20 kpc. 

Unfortunately, this method relies on the radial phase angle $\theta_r$, which in practice restricts its usage to axisymmetric potentials where the transformation to the action--angle space can be performed efficiently. Without using $\theta_r$, one could simply Fourier transform the 1-d energy distribution alone, which would still provide some measure of the impact done to the phase space, but would be far less sensitive. The overlapping of individual energy stripes, resulting from the fact that the energy stripes are at an angle and not exactly parallel, causes a smearing of the 1-d energy distributions. Any separation, or blurring by subhalo, between the energy stripes is therefore much clearer in the 2-d distributions -- see Figure~\ref{fig:1d2d} for an example.

\section{Simulation Setup}\label{simulations}

In this section, we describe the specific details of the various simulations conducted for this work. This includes the following:

\begin{itemize}
    \item an $N$-body simulation of a GSE-like merger, used to disperse the merger debris and to construct an approximate axisymmetric potential of the host galaxy for the subsequent test-particle integrations;
    \item a series of test-particle simulations, conducted within this potential, to assess the impact of general perturbing subhaloes on the GSE-like debris;
    \item simulations of the approximate trajectories of the LMC \& Sgr, which include the reflex motion of the host, to assess their impact on the GSE-like debris.
\end{itemize}

\subsection{Simulating a GSE-like Merger}\label{simingthesausage}

\begin{figure}
    \centering
    \includegraphics[width=0.48\textwidth]{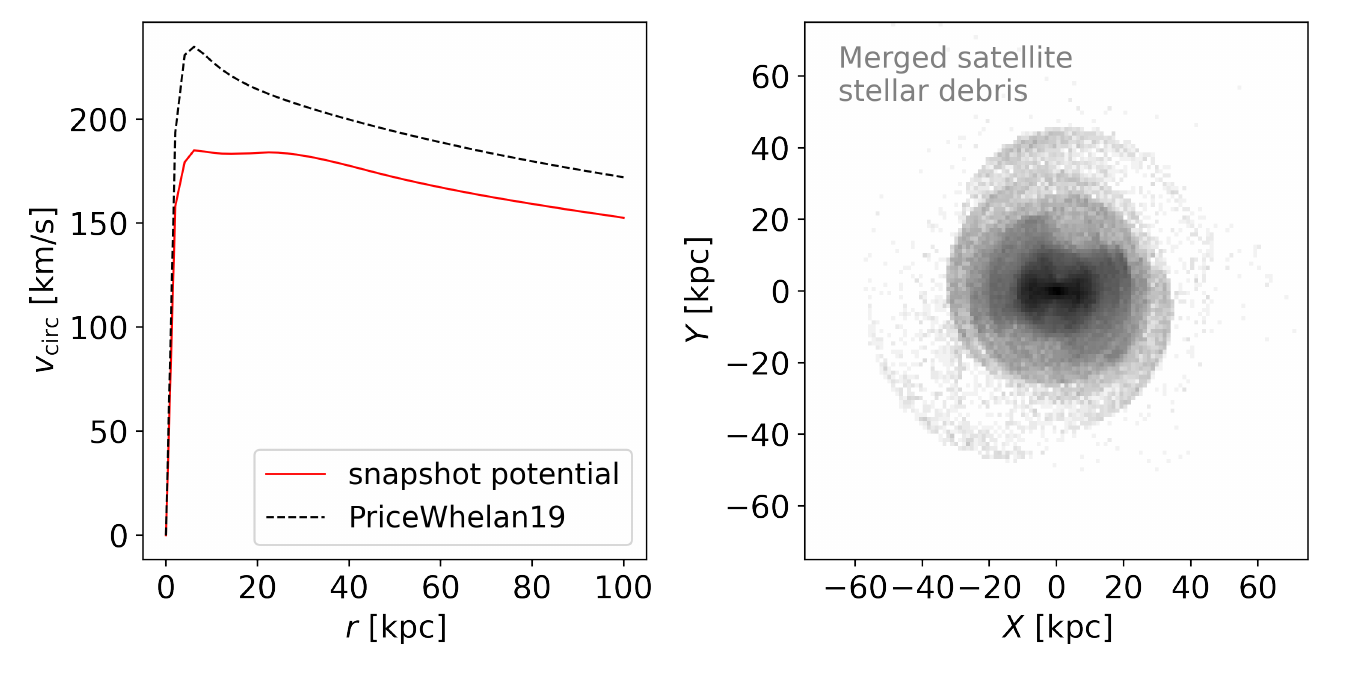}
    \caption{The final snapshot of the {\it N}-body simulation. {\it Left:} Rotation curve of the total (host \& satellite) multipole potential produced from the final snapshot ($t = 5$ Gyr) of the simulated satellite and host after merging, compared with the rotation curve of \textsc{MilkyWayPotential} potential from \textsc{Gala}. {\it Right}: Face-on view of the satellite stellar debris in configuration space.}
    \label{fig:rot_curve}
\end{figure}

To reproduce a GSE-like merger, we follow a similar setup to \citet{naidu2021reconstructing} and use their best fit initial conditions as a guide for our simulation, as presented in \citet{belokurov2022energy}. Specifically, we use a 1:2.5 total mass merger, with a satellite stellar mass of $M_* = 5 \times 10^8$~M$_{\odot}$ and a DM mass of $M_{\rm DM} = 2\times10^{11}$~M$_{\odot}$, placed on a prograde orbit with an inclination of 15$^{\circ}$ and a circularity of $\eta = 0.5$, where $\eta = L/L_{\rm circ}(E)$ is the ratio of total angular momentum to the angular momentum of a circular orbit of the same energy energy $E$. The  mass of the host halo is $M = 5\times 10^{11}$~M$_{\odot}$. At a lookback time of 5 Gyr, the result is a weakly triaxial merged host--satellite system, with axis ratios $1:0.87:0.33$ (which we approximate as an axisymmetric one for the purpose of computing angles and test particle integration), and a mass enclosed within $100$ kpc equal to about $4/5$ that of the \textsc{MilkyWayPotential} potential in \textsc{Gala} \citep[][]{price-whelan2017gala}. A density plot of the final snapshot, alongside the rotation curve, can be seen in Figure~\ref{fig:rot_curve}. The (total) mass enclosed within 10 kpc is $M_{\rm host}(r \leq 10 \: {\rm kpc}) = 0.8 \times 10^{11}$ M$_{\odot}$, the mass enclosed within 50 kpc is $M_{\rm host}(r \leq 50 \: {\rm kpc}) = 3.4 \times 10^{11}$ M$_{\odot}$, and the mass enclosed within 100 kpc is $M_{\rm host}(r \leq 100 \: {\rm kpc}) = 5.4 \times 10^{11}$ M$_{\odot}$. The initial $N$-body models are constructed using the \textsc{Agama} framework. We evolve the simulation from a look-back time of $t_{\rm lb} = 10$ Gyr up to $t_{\rm lb} = 5$ Gyr using the python package \textsc{PyFalcon}, a stripped down python interface of the \textsc{GyrFalcon} code \citep[][]{dehnen2000celestial}. The timestep was chosen as $\delta t = 2^{-10}$ Gyr, giving 5120 total steps.

The initial satellite consists of a dark halo component modelled by a truncated NFW profile
\begin{equation}\label{eq:nfw}
    \rho_{\rm NFW}(r) = \frac{\delta_c\rho_c\exp(-r / r_{\rm cut})}{(r /r_s) \left(1 + r / r_s\right)^2},
\end{equation}
with a scale radius of $r_s = 12$ kpc and an outer cut-off radius of $r_{\rm cut} = 60$ kpc, and a stellar component modelled by a Sersic profile with scale radius of $2$ kpc. In the above equation, $\rho_c \simeq 130$ M$_{\odot}$ kpc$^{-3}$ is the critical density of the universe and $\delta_c$ is the typical overdensity parameter for $c_{\rm 200}$.
The concentration parameter is defined by $c_{200} = r_{200}/r_s$, where $r_{200}$ is the radius within which the mean density is $200 \rho_c$. The satellite galaxy is represented by $2\times10^5$ particles assigned as stars and $10^6$ total particles, where the star particles are chosen since they are the most bound to their progenitor. 
The initial host consists of a halo component modelled also by a truncated NFW profile, with scale radius $15$ kpc and outer cut-off radius $120$ kpc, a bulge component modelled by a Sersic profile with scale radius $0.8$ kpc, and a disk component modelled by an exponential profile with scale radius $2$ kpc. The host has twice as many particles as the satellite.

After evolving the {\it N}-body simulation for 5 Gyr, we create a static axisymmetric potential from the final {\it N}-body snapshot using a multipole expansion, and save the final positions and velocities of the satellites $2\times10^5$ stellar debris particles. From a look-back time of $t_{\rm lb} = 5$ Gyr to the present, we represent the stellar debris as test particles in this static multipole approximation, as in \citet[][]{han2022tilt}, assuming that the self-gravity of the satellite is no longer relevant since all of the satellite particles are now bound to the host. We move from an {\it N}-body to a test particle simulation to significantly shorten the computation time, which is more substantial especially when considering a large number of subhaloes. One immediate limitation is that this reduced mass Milky Way will result in a somewhat inaccurate representation of the current behaviour of the LMC and Sgr. To combat this, we scale down the radii and mass of these satellites. However, for an experiment concerning dark matter subhaloes in the earlier history of the Milky Way (only shortly after the GSE has fully merged), this reduced mass potential will provide a reasonable backdrop in which to explore the effects of the dark matter subhaloes on the phase space substructure.

\subsection{Simulating the Generic Perturbing Subhaloes}

When including perturbing subhaloes, we consider two distinct scenarios: a) the impact of individual subhaloes and b) the impact of numerous populations of subhaloes, whose mass and number are determined by the subhalo mass function (the number density of haloes of different mass) of form \citep[e.g.][]{springel2008aquarius}, 
\begin{equation}\label{eq:SHMF}
    \frac{dN}{dM} = a_0\left(\frac{M}{m_0}\right)^n,
\end{equation}
for a mass range of $M=3.24\times10^{4}$ M$_{\odot}$ up to about $M = 2\times10^{10}$ M$_{\odot}$. We assume the subhalo mass function with constants $a_0,\: m_0$ corresponding to the 'A' halo in Aquarius simulations, that is $n=-1.9$, $a_0 = 3.26 \times 10^{-5}$ M$_{\odot}^{-1}$, and $m_0 = 2.52 \times 10^7$ M$_{\odot}$ \citep[][]{springel2008aquarius}. 

While the first $5$ Gyr are a {\it N}-body simulation, as described in Section \ref{simingthesausage}, in the final $5$ Gyr the merged satellite debris is represented as test particles in a time dependent potential. The new time dependent potential consists of the static multipole host potential and a collection of moving \citet[][]{hernquist1990analytical} potentials representing perturbing subhaloes:
\begin{equation}
    \rho_{\rm sh}(r) = \frac{M}{2\pi} \frac{a}{r}\frac{1}{(r+a)^3}.
\end{equation}

The case for using this potential for the subhaloes is explained in \citet[][]{springel2005modelling} and, as in their work, we match this Hernquist profile with a corresponding NFW potential \citep[][]{navarro1996structure} of the same virial mass. The virial radius is related to the mass by $r_{200} = 210 \times (M / 10^{12}$~M$_{\odot})^{1/3}$. For a given NFW mass, we obtain an estimate for the concentration $c=r_{200}/r_s$ by the concentration-mass relation \citep[e.g.][]{ludlow2016mass, gilman2019probing}. By enforcing equal inner density profiles, we obtain a relationship between the scale radius of the Hernquist potential $a$, and the NFW scale radius $r_s$:
\begin{equation}
    a = r_s \sqrt{2\left[\ln(1+c) - c/(1+c)\right]}.
\end{equation}

The moving subhaloes are initialised as test particles, and integrated in the static multipole potential, without the satellite debris. We then construct a new composite time-dependent potential consisting of the static host galaxy potential plus a collection of moving Hernquist potentials centred on the pre-computed subhalo trajectories. It is this composite potential in which the satellite debris is represented as massless tracer particles. By doing so, we ignore any reflex motion that these subhaloes may induce on the host galaxy, which is indeed negligible for subhalo masses below $10^{10}\,M_\odot$. We justify this assumption by the fact that the mass of a single subhalo with $M \lesssim 10^{10} M_{\odot}$, has only a fraction of the mass of the MW enclosed within its orbit.

\subsubsection{Single subhalo method}

For experiments investigating the impact of single subhaloes, we consider masses of $M=10^6, 10^7, 10^8, 10^9$ and $10^{10}$ M$_{\odot}$. For a given mass, we also explore how the pericentre affects the results; the subhaloes orbits are set up such that they have a pericentre of either $r_{\rm peri} = 10, 30$ or $50$ kpc. In all cases, the subhalo apocentre is set at $100$ kpc, and each orbit is confined entirely to the equatorial plane so that their impact on the debris is maximal. We take a concentration of $c=18, 16, 14, 12$ and $10$ for the respective masses in increasing order. Beyond changing the properties of the subhalo, we also consider different orbital periods. We quantify the impact on the satellite debris in two scenarios: a) after the subhalo makes a single fly-by from apocentre to apocentre, and b) after the subhalo makes multiple fly-bys, continually perturbing the debris for 5 Gyr. In both scenarios $A_S$ is calculated after a total integration time of 10 Gyr, so that we always compare the ironing of the final snapshot. Therefore, for single flybys, the subhaloes are introduced later in the simulation.

\subsubsection{Multiple subhalo method}

After inspecting the impact of single subhaloes, it seems reasonable to examine the case of many subhaloes perturbing at once. This explores a scenario more like the genuine history of the MW, undergoing numerous simultaneous perturbing events at once. Therefore, we present a series of simulations with varying numbers of subhaloes, $N_{\rm sh}$, in accordance with the subhalo mass function. We consider only the four largest masses of the previous experiment, and integrate equation \ref{eq:SHMF} to obtain a subhalo number. Specifically, we integrate over a $\log_{10}$ mass range of 0.4 around these four masses,
\begin{equation}\label{eq:integrateshmf}
    N_{\rm sh}(M) = \int_{M\times 10^{-0.2}}^{M\times 10^{+0.2}} \left(\frac{dN}{dM'}\right) dM', 
\end{equation}
and obtain the numbers found in Table~\ref{tab:mass_number_table}. These subhalo numbers are in reasonable agreement with the expected number of subhaloes from other zoom-in simulations \citep[e.g.][]{diemand2007formation, nadler2022symphony}.

\begin{table}
\centering
\caption{The number of subhaloes of each mass in the multiple subhalo experiments. While all subhaloes have the same mass $M$, in a given experiment, the number of subhaloes is determined by integrating the subhalo mass function over a range around $M$: $\log_{10} M \pm 0.2$.}
\begin{tabular}{|l|l|}
\hline
Subhalo Mass, $M$ {[}M$_{\odot}${]} & $N_{\rm sh}(|\log_{10} M| \leq 0.2)$ \\ \hline
$10^{10}$                           & 4                                 \\ \hline
$10^9$                              & 28                                \\ \hline
$10^8$                              & 225                               \\ \hline
$10^7$                              & 1789                              \\ \hline
\end{tabular}
\label{tab:mass_number_table}
\end{table}

As with the previous experiments, the simulations are run for a total of $10$ Gyr, where the first $5$ Gyr are a {\it N}-body merger simulation, and in the final $5$ Gyr the satellite debris is represented as test particles in a time dependent potential. We sample the initial positions and velocities of the subhaloes from a density and distribution function that very roughly approximates that of the known MW satellites. The density is a modelled by an \textsc{Agama} \textsc{Spheroid} profile with $(\alpha, \beta, \gamma) = (2.0, 6.0, 0.5)$ and scale radius 100 kpc, and the corresponding isotropic \textsc{QuasiSpherical} distribution function is constructed using the Eddington inversion formula. A selection function is also applied to force the subhaloes to begin with an inward radial velocity, and at a distance of $70$ kpc $< R < 150$ kpc from the centre of the host potential, so that they are far enough from the debris to not have an instantaneous effect on the debris.  
The selection function also enforces a maximum pericentre $r_{\rm peri}$ on the subhaloes, so that we may examine the relationship between $(v_r,r)$ substructure disruption and subhalo pericentre just like the single subhalo experiments. For a given $N_{\rm sh}$, we conduct two simulation that enforce a maximum pericentre of $r_{\rm peri} = 20$ kpc and $r_{\rm peri} = 50$ kpc, respectively.

\subsection{Simulating the LMC \& Sagittarius}

When considering perturbations of the LMC and Sgr, we use a more sophisticated method than simply placing a Hernquist potential onto a moving test particle. We follow the method described in greater detail in \citet[][]{vasiliev2021}, whereby the reflex motion of the host toward the LMC is accounted for. Note that we account for the reduced mass of the resulting multipole potential by reducing the mass of the LMC and Sgr by 4/5, and by scaling the respective scale radii and outer-cutoff radii appropriately.

The LMC contributes two potential components: a direct gravitional potential and a uniform time-dependent acceleration resulting from the reflex of the host towards the LMC. The LMC is modelled by a truncated NFW, with total mass $M_{\rm LMC} = (4/5) \times 1.4 \times 10^{11}$ M$_{\odot}$, 4/5 the mass of the LMC found in \citet[][]{erkal2019total}, and a scale radius and outer-cutoff radius of $r_{\rm s} \sim 9$ kpc and $r_{\rm cut} = 100r_s$, repectively. Since the LMC has only just passed its first pericentre of $\sim 50$ kpc, it is appropriate to approximate the the LMC and MW as two rigid mutually attracting gravitating galaxies without concerning oneself with the internal deformations (though see \citealt{lilleengen2022} for an investigation of impact of MW and LMC deformations). The trajectory of the MW and LMC is described by the following system of coupled differential equations:

\begin{align}\label{eq:lmcdiffeq}
\begin{split}
    \dot{\bm{x}}_{\rm LMC} &= \bm{v}_{\rm MW}, \\
    \dot{\bm{v}}_{\rm MW} &= -\nabla \Phi_{\rm LMC} (\bm{x}_{\rm MW} - \bm{x}_{\rm LMC}), \\
    \dot{\bm{x}}_{\rm LMC} &= \bm{v}_{\rm LMC}, \\
    \dot{\bm{v}}_{\rm LMC} &= -\nabla\Phi_{\rm MW} (\bm{x}_{\rm LMC} - \bm{x}_{\rm MW}) + \bm{a}_{\rm DF},
\end{split}
\end{align}
where $\Phi_{\rm MW}$ and $\Phi_{\rm LMC}$ are the static potentials of the respective galaxies. The parameter $\bm{a}_{\rm DF}$ is the Chandrasekhar dynamical friction acceleration, defined by 

\begin{equation}
    \bm{a}_{\rm DF} = \frac{-4\pi \rho_{\rm MW} G^2 M_{\rm LMC} \ln \Lambda}{v^2}\left[\text{erf}(X) - \frac{2X \exp(-X^2)}{\sqrt{\pi}} \right]\frac{\bm{v}}{v}
\end{equation}
where $X = v/\sqrt{2}\sigma_{\rm MW}$. In the above equations, $\rho_{\rm MW}$ and $\sigma_{\rm MW}$ are the density and velocity dispersion of the host MW potential, and $\ln \Lambda$ is the Coulomb logarithm. Equations~(\ref{eq:lmcdiffeq}) are integrated backwards in time, from the present, for 5 Gyr. The result is used to used to compute the uniform acceleration potential that the satellite stellar debris is integrated in, for 5 Gyr up to the present.

We also consider the influence of Sgr, accompanied by the LMC. To generate a realistic model of Sgr with decaying mass, we adopt the method described in detail by \citet{dillamore2022impact}. Briefly, the Sgr model loses a fixed fraction of its total mass at each pericentre passage until it reaches a present-day value of $(4/5) \times 4\times10^8$ M$_\odot$, which is 4/5 the value of that found in \citet{vasiliev2020last}. Its orbital trajectory (shown in Figure~\ref{fig:lmc_sgr_traj}) is consistent with dynamical friction acting on a particle with this mass decay profile. We consider a $\delta=1.0$ model of \citet{dillamore2022impact} which, in our potential, has mass $2\times10^{10}M_\odot$ at infall and loses $90\%$ of its mass with each pericentre passage. There are multiple lines of evidence for an initial mass of this value or larger \citep[e.g.][]{gibbons2017sgr, read2019abundance, laporte2019sgr, bennett2022exploring}. We include the effect of the reflex motion of the MW's centre towards Sgr via the same method as the LMC.

\section{Impact of Subhaloes}\label{genericsubhaloes}

\begin{figure*}
    \centering
    \includegraphics[width=0.98\textwidth]{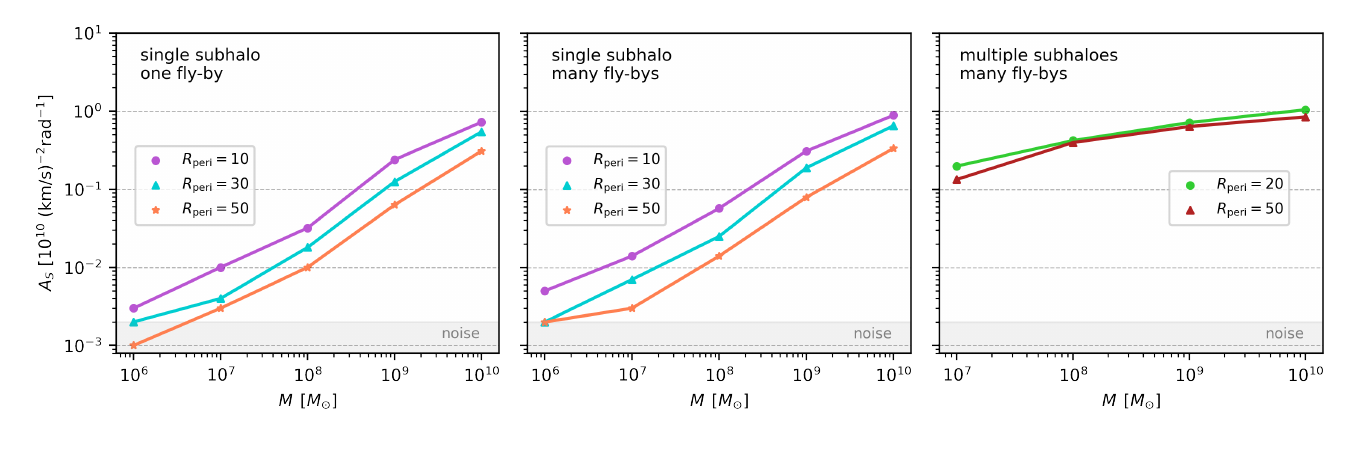}
    \caption{The value of the ironing parameter $A_S$ for each of the three generic subhalo experiments, which explore the dependence of $A_S$ on mass, pericentre and number of subhaloes. In all cases $A_S$ is calculated from the test particle simulation, following the initial {\it N}-body simulation. The grey shaded area indicate the noise, which is $\sigma_A~=~0.002$~(km/s)$^{-2}$rad$^{-1}$. {\it Left:} The value of $A_S$ after a single fly-by of a single subhalo, for time of a single orbital period following the initial {\it N}-body simulation. {\it Middle:} The results after the debris undergoes continuous perturbation by the subhalo for 5 Gyr. {\it Right:} The results after continuous perturbations by multiple subhaloes for the 5 Gyr.}
    \label{fig:all_lineplot}
\end{figure*}

In this section, we present the results of a series of simulations of the perturbed GSE-like merger which are run for a total of $10$ Gyr, where the first $5$ Gyr are a {\it N}-body simulation as described in Section \ref{simingthesausage}. In the final $5$ Gyr, the merged satellite debris is represented as test particles in a time dependent potential which consists of static multipole host potential combined with a collection of moving Hernquist subhalo potentials. We introduce subhalo perturbers in order to see how the satellite debris responds to perturbations in $(v_r, r)$ and $(E, \theta_r)$ coordinates. We describe the results of four distinct experiments: 

\begin{enumerate}
    \item the impact of a single subhalo which makes only a single fly-by $(T \sim 1$ Gyr) from apocentre to apocentre
    \item the impact of a single subhalo which makes multiple fly-bys, and therefore continually perturbs the debris for 5 Gyr.
    \item the impact of many subhaloes, which make multiple fly-bys and continually perturb the debris for 5 Gyr.
    \item The impact of subhaloes which approximate the recent behaviour of known MW satellites, specfically the LMC and Sgr.
\end{enumerate}

For the first three experiments, we assess how $A_S$ depends on pericentre, which is described in more detail in each sub-section. We calculate $A_S$ and compare it with the noise $\sigma_A$ at relevant time steps to get a preliminary measure of the impact done by the perturbing halo. In certain cases, we provide illustrative examples of the impact on $(v_r, r)$ and $(E, \theta_r)$ space. All of the values of $A_S$, for the first three experiments, are displayed in Figure~\ref{fig:all_lineplot}, where we see an approximate power law trend for the single subhalo experiments, for the range of masses that we consider. However, one would expect that the trend flattens out at either end as the disturbance to the stripes either reaches the level of noise, or reaches the level of maximum disturbance -- where the stripe distribution turns uniform. 

\begin{figure}
    \centering
    \includegraphics[width=0.48\textwidth]{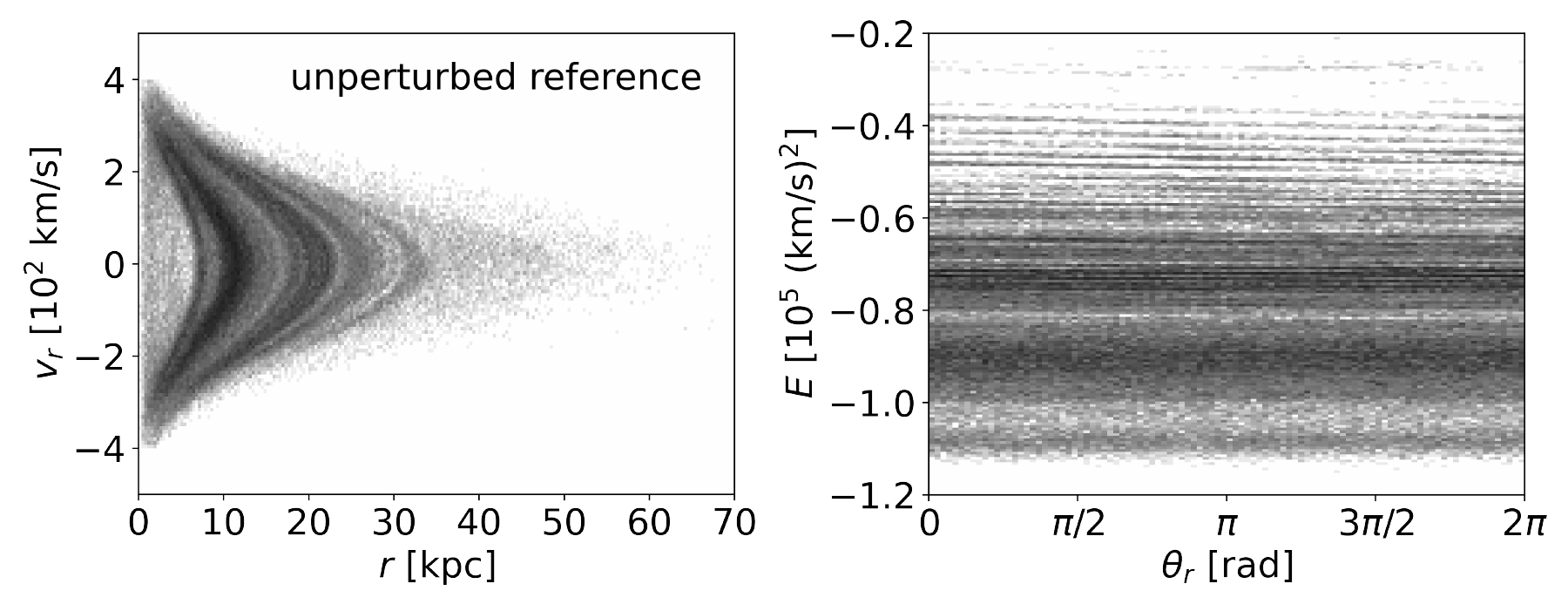}
    \caption{The final snapshot of the evolution of the $(v_r, r)$ and $(E,\theta_r)$ space after 5 Gyr of unperturbed evolution. This figure is to be used as a reference for comparison with subsequent figures.}
    \label{fig:reference}
\end{figure}

See Figure~\ref{fig:reference} for a visual comparison with the unperturbed case, where we show the $(v_r,r)$ and $(E,\theta_r)$ snapshot after 5 Gyr with no perturbing subhalo.

\subsection{Single-subhalo interactions}

Here, we present the results of simulations with only a single subhalo perturbing the satellite debris. We consider subhaloes with masses of $M=10^6, 10^7, 10^8, 10^9$ and $10^{10}$ M$_{\odot}$. For a given mass, we also explore how pericentre affects the results.

\subsubsection{Single fly-by}\label{sec:singleflyby}

\begin{figure}
    \centering
    \includegraphics[width=0.47\textwidth]{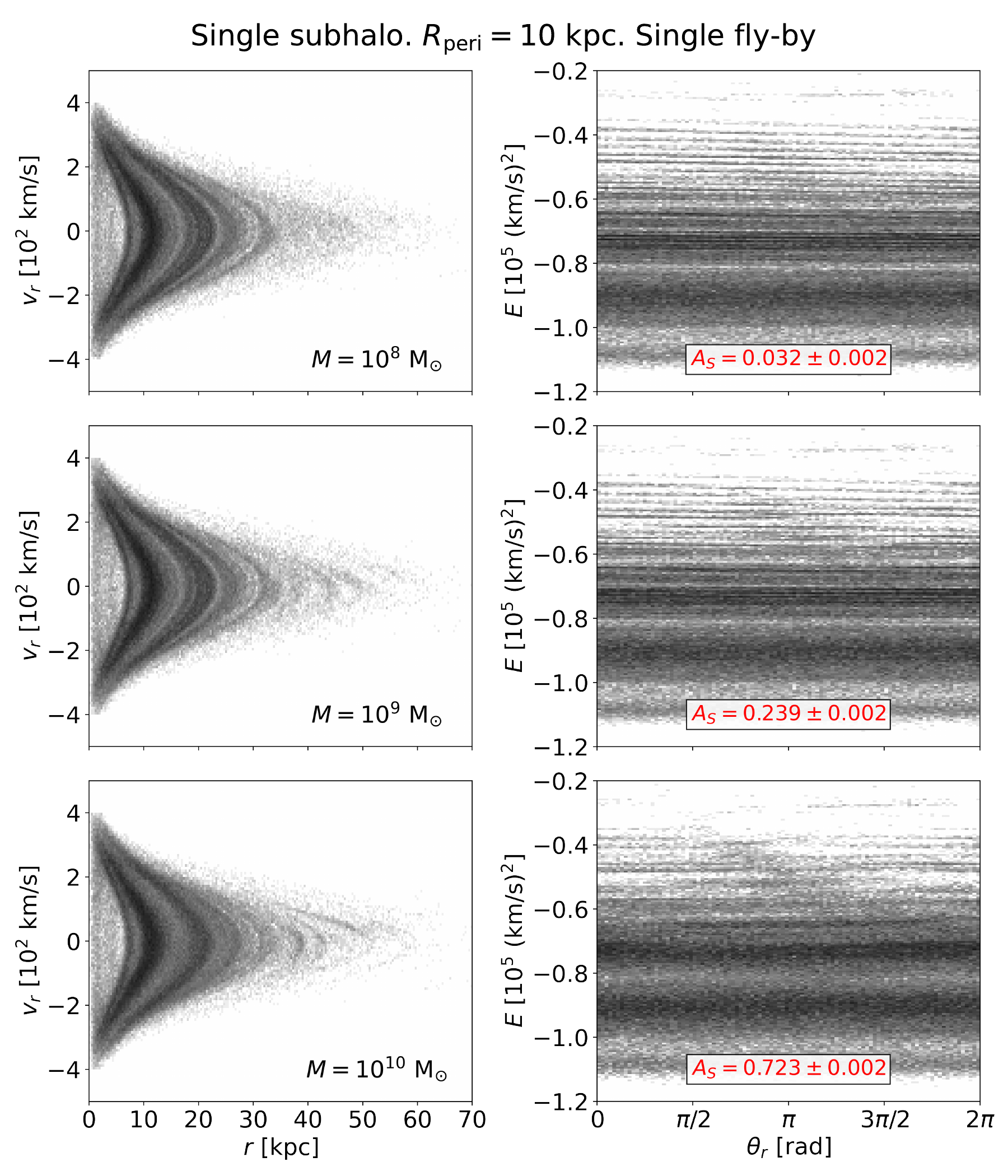}
    \caption{Impact of single perturbing subhaloes with pericentre of 10 kpc, for a variety of masses, after the subhalo completes one orbital period. The snapshot, and the value of $A_S$, shown are at 5 Gyr after the end of initial {\it N}-body simulation. The units of $A_S$ are (km/s)$^{-2}$rad$^{-1}$. {\it Left column:} $(v_r, r)$ space of the satellite debris, for each mass subhalo. {\it Right column:} $(E, \theta_r)$ space for the corresponding panel to the left. The value of $A_S$ for each snapshot is presented in red text. The bottom panel appears to show the path of the subhalo through energy-angle space.}
    \label{fig:single_flyby}
\end{figure}

For single fly-by experiments, we calculate $A_S$ at the approximate time that a subhalo makes one single orbit from apocentre to apocentre. Since different pericentres correspond to different orbital periods, we ensure that the subhaloes make a single fly-by, such that their orbit is completed at the instant of final timestep of the 5 Gyr test-particle integration. 

The results of the single subhalo interaction experiments are shown in the left panel of Figure~\ref{fig:all_lineplot}. Additionally, we visualise the radial phase space and energy-angle space of the $r_{\rm peri} = 10$ kpc experiment in Figure~\ref{fig:single_flyby}. Immediately evident is that for a fixed mass, increasing the pericentre increases the ironing. Moreover, for a fixed pericentre, the ironing is increased as mass increases. For all pericentres, subhaloes with mass $M=10^6$ M$_{\odot}$ do very little damage to the satellite's $(E, \theta_r)$ space, all with $A_S \sim \sigma_A$. Moreover, even for $M=10^8$ M$_{\odot}$, the impact remains low with $A_S \lesssim 0.05$. For comparison, by examining the top panel of Figure~\ref{fig:single_flyby} we can see that for $A_S = 0.037$ numerous chevrons remain visible, and the line structure is virtually unchanged. The most destructive example of the single fly-by experiment is presented in the bottom panel of Figure~\ref{fig:single_flyby}. In this case, with $A_S = 0.723$, the impact done to the $(E, \theta_r)$ stripes is more substantial. Correspondingly, many of the $(v_r, r)$ chevrons have been smoothed out; there is less substructure present at $r < 30$ kpc than in the lower mass cases. There is a notable feature in the bottom panel: a blurred streak at an $\approx 45^\circ$ angle through the high energy ($E > -0.6$ (km/s)$^2$) portion of the energy-angle distribution. 

While chevrons can be destroyed or smoothed, sometimes fragments of chevrons appear to be produced by the perturber. Most notably in the case of the highest mass perturber, we see the appearance of chevrons at higher radii at around $r\sim 40$ kpc, which are not present in the  unperturbed simulations. By inspection of the associated $(E, \theta_r)$ space, one can see that the perturber has created new energy clumps at high energies which corresponding to the appearance of new fragments of chevrons. This effect is reproduced in later experiments.

\subsubsection{Multiple fly-bys}

\begin{figure}
    \centering
    \includegraphics[width=0.47\textwidth]{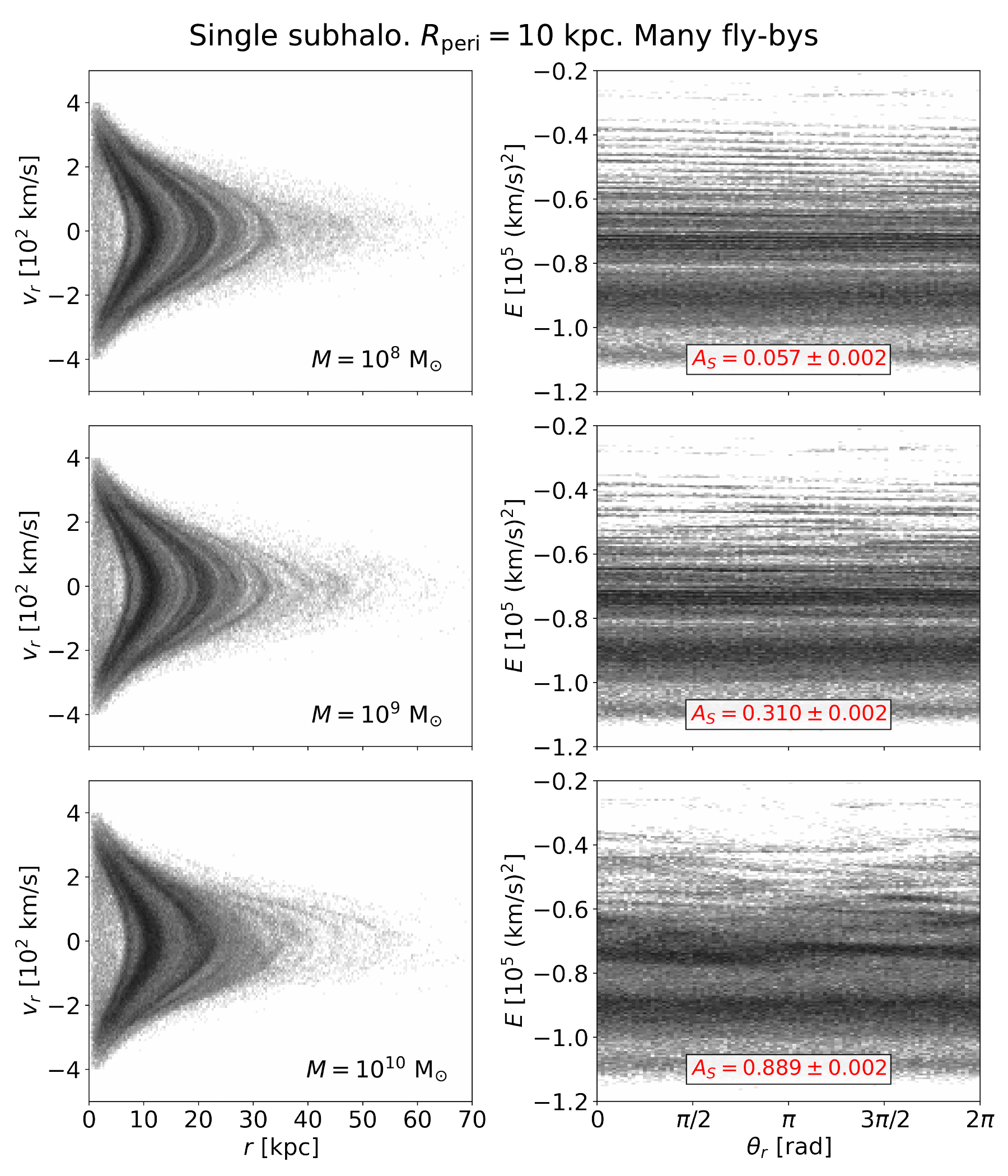}    \caption{Impact of single perturbing subhaloes with pericentre of 10 kpc for a variety of masses. The snapshot, and the value of $A_S$, shown are at 5 Gyr after the end of initial {\it N}-body simulation. The units of $A_S$ are (km/s)$^{-2}$rad$^{-1}$. {\it Left column:} $(v_r, r)$ space of the satellite debris, for each mass subhalo. The impact of the subhalo is most notable in the bottom panel, where some of the chevrons have been smoothed out. {\it Right column:} $(E, \theta_r)$ space for the corresponding panel to the left. The value of $A_S$ for each snapshot is presented in red text. The bottom panel illustrates the most destructive nature of the heavier subhalo.}
    \label{fig:peri10_allmass_t5}
\end{figure}

\begin{figure}
    \centering    \includegraphics[width=0.47\textwidth]{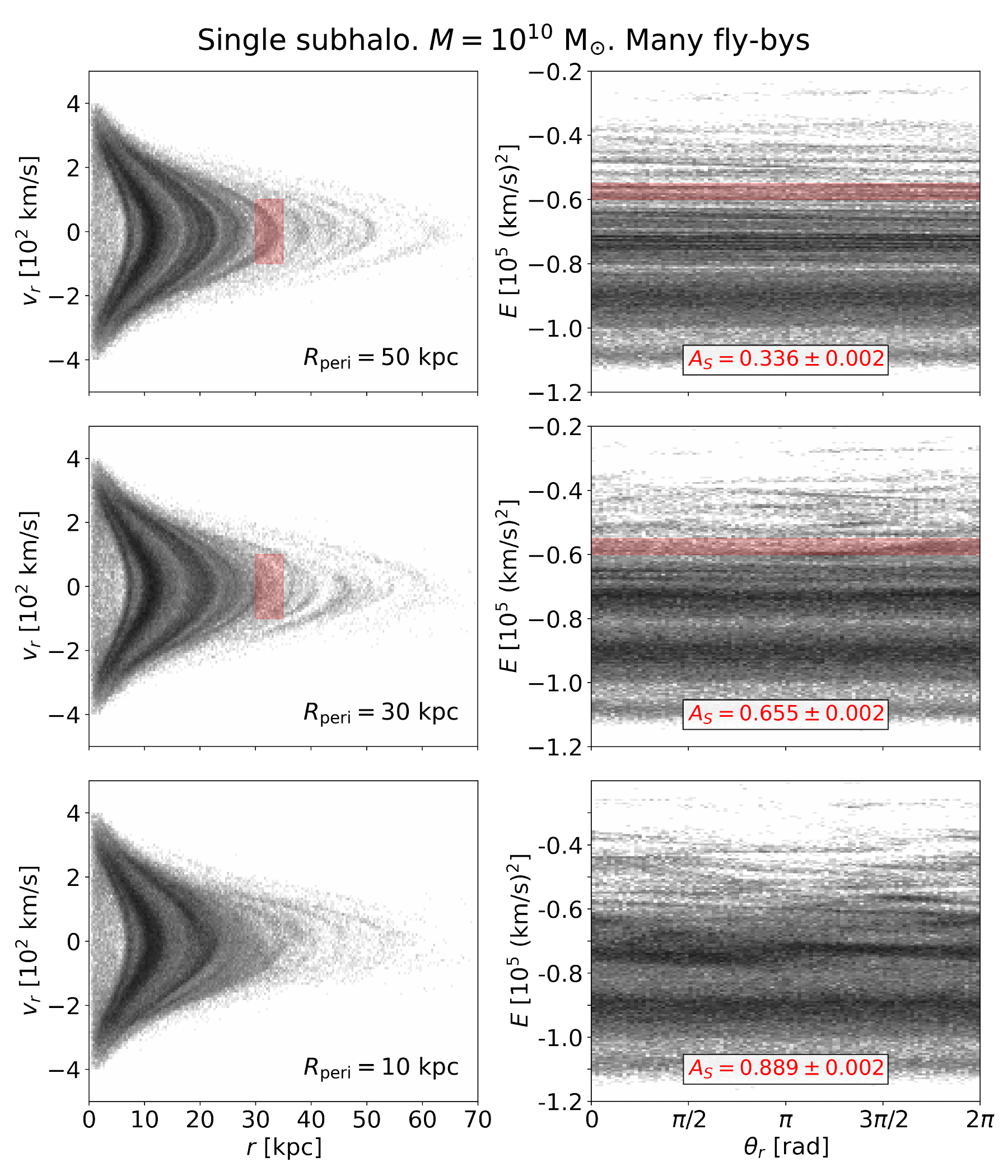}    \caption{Impact of single perturbing subhaloes with a mass of $10^{10}$ M$_{\odot}$ for a variety of pericentres. The snapshot, and the value of $A_S$, shown are at 5 Gyr after the end of initial {\it N}-body simulation. The units of $A_S$ are (km/s)$^{-2}$rad$^{-1}$. {\it Left column:} $(v_r, r)$ space of the satellite debris, for each mass subhalo. The impact of the subhalo is most notable in the bottom panel, where some of the chevrons have been smoothed out. {\it Right column:} $(E, \theta_r)$ space for the corresponding panel to the left. The red shaded region marks out the destruction of a chevron when the pericentre of the subhalo is decreased.}
    \label{fig:mass10_allperi_t5}
\end{figure}

In this section, we present the results of similar simulations, except that we calculate $A_S$ after continuous perturbation by the subhaloes for 5 Gyr. In this scenario, the subhaloes in all cases make multiple fly-bys through the debris. The results of this experiment are presented in the middle panel of Figure~\ref{fig:all_lineplot}. Figure~\ref{fig:peri10_allmass_t5} illustrates the result of varying subhalo mass, for a fixed pericentre, whilst Figure~\ref{fig:mass10_allperi_t5} illustrates the impact of varying pericentre for a fixed mass. In both Figures, we neglect to include the results for the two lightest mass subhaloes as their impact is weaker than that of the $10^8$ M$_{\odot}$ subhalo, whose visible impact on phase space is already minimal.

Comparison of the left and middle panels of Figure~\ref{fig:all_lineplot} show that, in all cases, multiple fly-bys cause more impact than single fly-bys. Figure~\ref{fig:peri10_allmass_t5} show that subhaloes with $M \leq 10^8$ M$_{\odot}$ remain ineffective at disrupting the phase space substructure, with $A_S = 0.057$ for the lowest pericentre value and $A_S < 0.057$ for all other pericentres. For $r_{\rm peri} = 50$ kpc, the change in subhalo impact from single fly-by to many fly-bys was quite minimal, whereas for smaller pericentres, the ironing increased by up to a factor of about 1.5. 

Figure~\ref{fig:peri10_allmass_t5} shows that even for very eccentric subhalo orbits with $r_{\rm peri}=10$ kpc, only the $10^{10}$ M$_{\odot}$ subhalo is able to severely disrupt chevrons at $r\sim20$--$30$ kpc with $A_S = 0.889$. Some impact is noticeable at $r\sim30$ kpc chevron for the $10^9$ M$_{\odot}$ subhalo, with $A_S = 0.310$, while essentially no impact is done for the lowest mass subhalo, with $A_S = 0.057$. In the highest mass case, the chevron with tip at $r \sim 20$ kpc is found to be split into two halves. This effect is mirrored at $E \gtrsim -0.8$ (km/s)$^2$ in the energy-angle distribution.

Since we have established that, rather unsurprisingly, the  most massive subhalo is the most disruptive, it is worth visualising the behaviour of the satellite debris chevrons for this mass with a variety pericentres. In Figure~\ref{fig:mass10_allperi_t5}, we can see how the chevrons which are disrupted depend on the pericentre of the subhalo. In the top panel, $A_S = 0.336$ and while some energy stripes are changed in frequency, there remain visibly intact stripes. Examination of the corresponding $(v_r,r)$ coordinates shows chevrons at all values of $r$ remain clear and distinct. The middle panel illustrates the ironing of the vast majority of chevrons for $r>30$ kpc when the pericentre of the orbit is 30 kpc. In this case, the value of $A_S$ is about twice that of the panel above. We also mark in red the clear removal of a chevron at about $r \sim 40$ kpc which was present in the above panel, indicating the importance of the pericentre of the subhalo in destroying the innermost chevrons. Finally, in the bottom panel, we note that all chevrons with $r\gtrsim 10$ kpc are severely disrupted. The corresponding value of $A_S = 0.889$ is therefore indicative of severe disturbance to the phase space substructure.

The take-away from this experiment is that, broadly speaking, if a sufficiently high mass subhalo ($M \sim 0.1 \mathcal{O}(M_{\rm host})$)  perturbs phase-mixed satellite debris, then phase space chevrons with tips at $r_{\rm chev}$ will only be significantly disturbed if $r_{\rm peri} < r_{\rm chev}$. For a subhalo with comparatively low mass, even a pericentre of 10 kpc is not enough to disrupt the chevrons with $r_{\rm peri} \sim r_{\rm chev}$, but may still reduce the amplitude of the energy frequency distribution. This implies that even very large mass known MW satellites may not have severely disrupted the GSE phase space if they are sufficient far out. Yet sufficiently massive and eccentric subhaloes may leave their trace in phase space via the removal of outer chevrons. Should chevrons be found at high radii, this provides insight into the nature of MW subhaloes.

These experiments show that a value of $A_S \sim 1$ guarantees ironing of most of the chevron substructure. Intermediate values of $A_S \sim 0.5$ require further inspection, but are typically a sign of sign of changed substructure. However, ironing values of $A_S < 0.1$ are certain to have very little impact on the substructure. A low value of $A_S$ could indicate simply a shift in the amplitude of the frequency power spectrum.

\begin{figure*}
  \centering
  \subfigure{\includegraphics[scale=0.38]{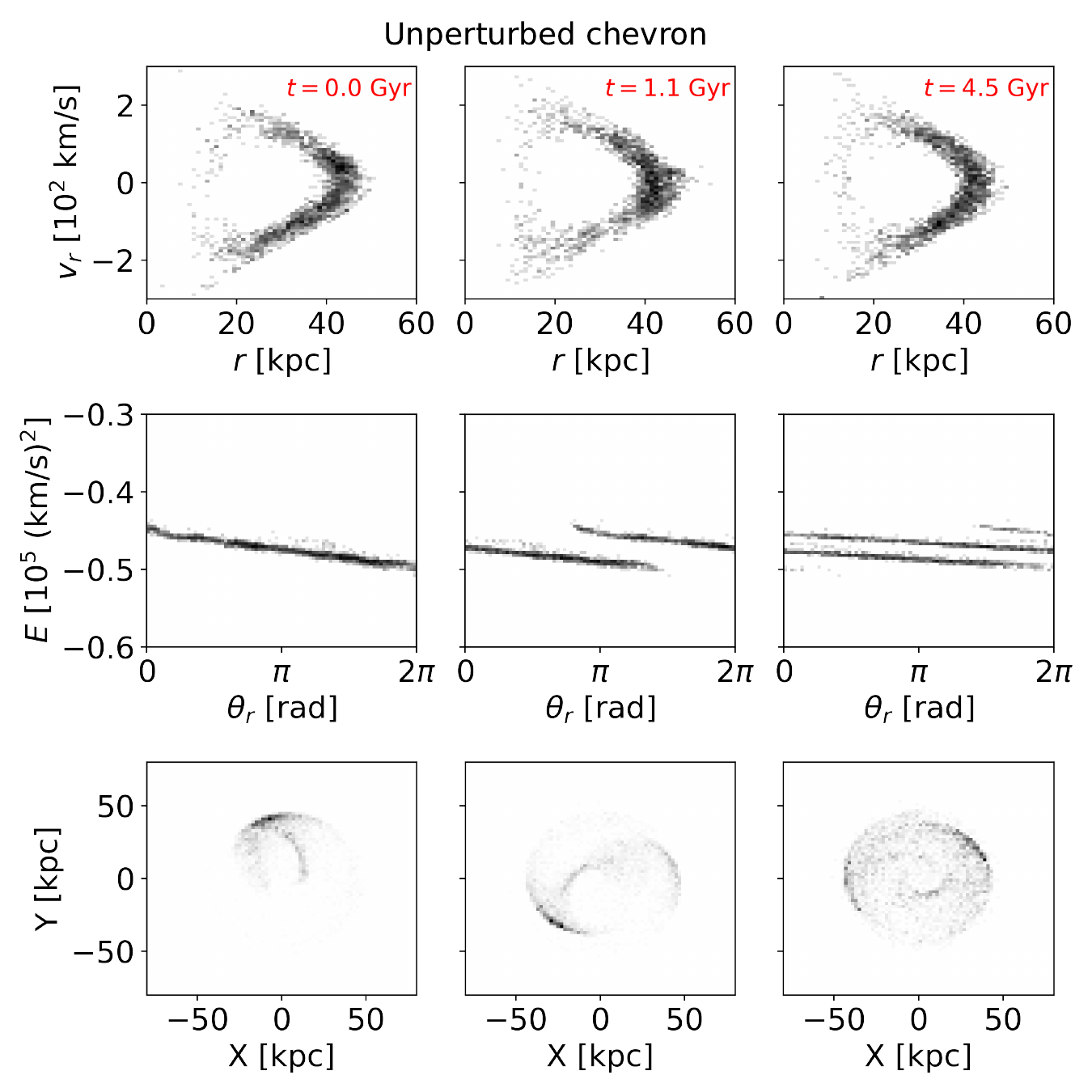}}\quad
  \subfigure{\includegraphics[scale=0.38]{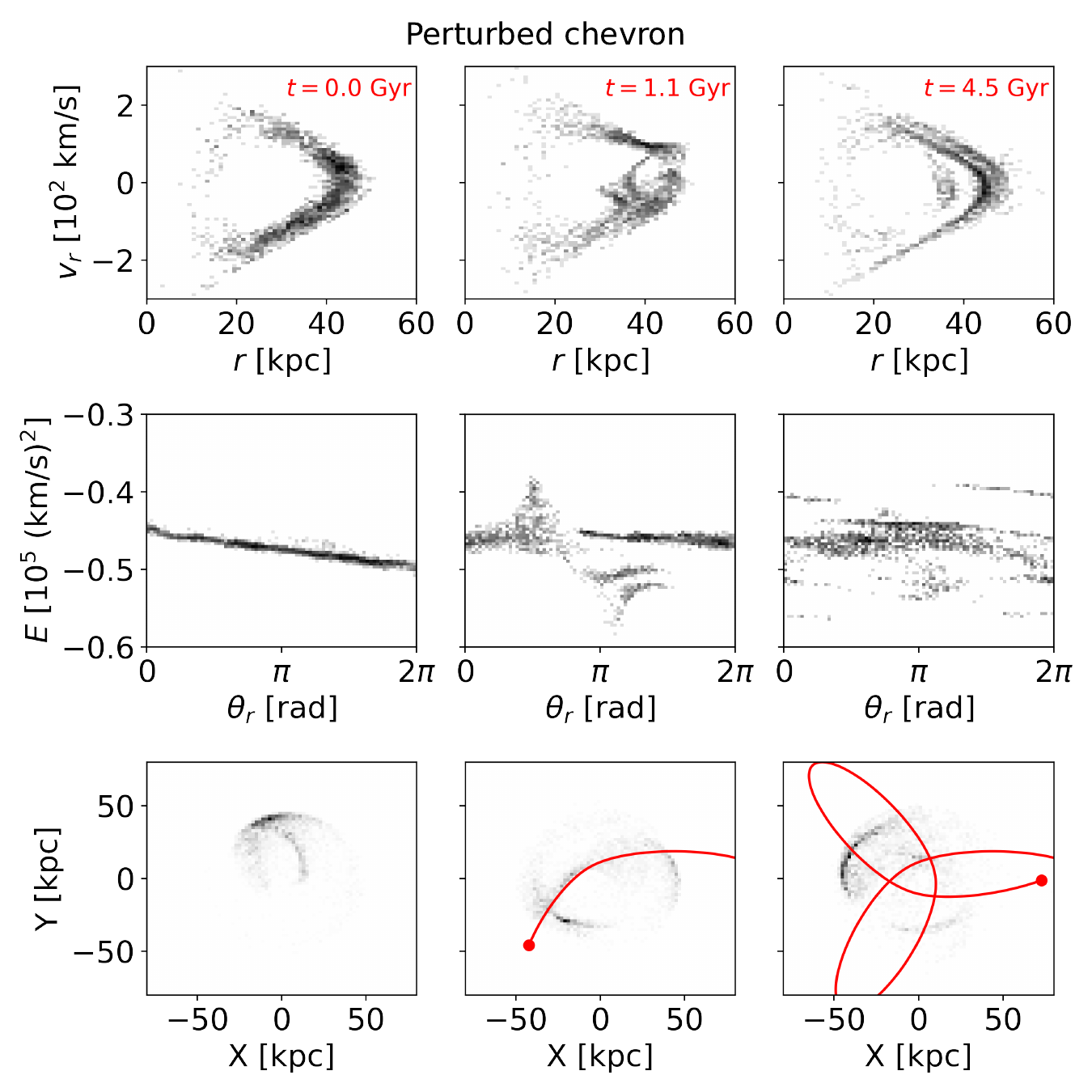}}
  \caption{Comparison of the evolution of a single phase space chevron with and without a perturbing subhalo. The subhalo has mass $M=10^{10}$ M$_{\odot}$ and pericentre of $10$ kpc. The single $(v_r,r)$ chevron is selected by choosing the particles in its associated energy stripe. {\it Left $3\times3$ panel:} Snapshots at 3 illustrative time-steps for the $(v_r, r)$, $(E,\theta_r)$ and $(x,y)$ spaces of the selected chevron, evolved in the static potential with no perturbing subhaloes. {\it Right $3\times3$ panel:} The same plots, at the same time-steps, but with a perturbing subhalo. The trajectory of the subhalo is indicated in the $(x,y)$ space by a red line and point. Note the disruption done to the energy-angle space after 4.5 Gyrs of perturbing and corresponding fragmented chevron in radial phase space.}
  \label{fig:individual}
\end{figure*}

Once again, as in the case of a high-mass single fly-by, the top two panel of Figure~\ref{fig:mass10_allperi_t5} (most clearly) show the creation of chevrons at high radii. We suggest that these ``artificial chevrons'' are the result of two processes. In the case of an especially high mass perturbing, some debris can be picked up and ``re-(phase-)mixed'' at high energies. However, more commonly particles have their energy more subtly changed by the perturbation causing debris from multiple stripes to overlap or bunch up and create these higher radii chevrons. In Figure~\ref{fig:individual}, we illustrate the creation of fragmented chevrons and bunching of energy by isolating a single $(v_r,r)$ chevron. The single chevron is isolated by drawing a polygon around it's stripe in $(E,\theta_r)$ space and selecting the particles contained within the polygon. We then follow only the selected particles for a subsequent 5 Gyrs and plot them in radial phase space, energy-angle space and configuration space. For comparison, we show the unperturbed chevron alongside. In the unperturbed case, the chevron continually evolves to form multiple stripes in $(E,\theta_r)$ space, as selecting an initially slanted energy line constitutes selecting particles with a variety of $\Omega_r(E)$. The corresponding $(v_r,r)$ space does not clearly show multiple chevrons as the energy range is narrow and the chevrons overlap. The behaviour in the perturbed case is quite different. Notable features to be seen in the rightmost column of Figure~\ref{fig:individual} include the fragmented chevron at lower radii, and the more subtle splitting of the largest chevron. The bunching of energies seen in the rightmost $(E, \theta_r)$ plot is a common feature seen in most highly perturbed experiments throughout this section, and typically explains the appearance of chevrons at radii where there were previously none visible. The range of energies is widened by the perturbing subhalo, and thus the chevrons change their appearance in $(v_r,r)$ space.

\subsection{Many-subhalo interactions}

\begin{figure}
    \centering
    \includegraphics[width=0.47\textwidth]{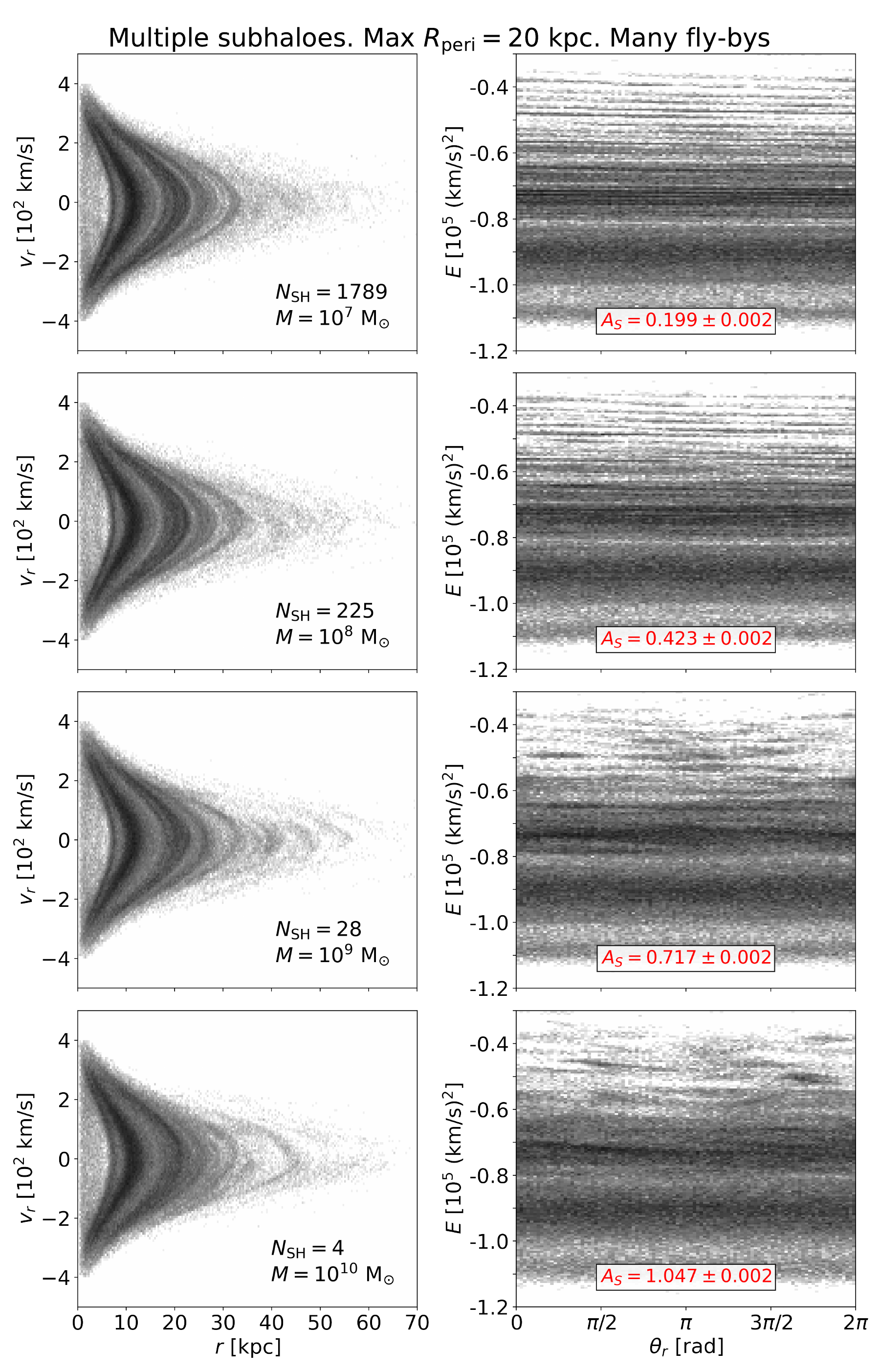}
    \caption{Impact of multiple ($N_{\rm sh} > 1$) perturbing subhaloes with a mass and number $N_{\rm sh}$ assigned in accordance with equation~\ref{eq:SHMF}. All subhaloes are sampled such that their pericentres are below 20 kpc. The snapshot, and the value of $A_S$, shown are at 5 Gyr after the end of initial {\it N}-body simulation. The units of $A_S$ are (km/s)$^{-2}$rad$^{-1}$. {\it Left column:} $(v_r, r)$ space of the satellite debris, for each mass subhalo. {\it Right column:} $(E, \theta_r)$ space for the corresponding panel to the left. The values of $A_S$ are significant for every mass-number combination.}
    \label{fig:many_subhalo_impact}
\end{figure}

In this section, we present the results of simulations with multiple subhaloes perturbing the satellite debris. We consider subhaloes with masses of $M=10^7, 10^8, 10^9$ and $10^{10}$ M$_{\odot}$, whose numbers are determined by equation \ref{eq:integrateshmf}. For a given mass-number combination, we also explore how an enforced maximum pericentre on the subhalo population affects the results.

The right panel of Figure~\ref{fig:all_lineplot} presents the ironing values for both maximum pericentre scenarios, with all four mass-number configurations. While the total cumulative mass in each case is somewhat similar, the values of $A_S$ vary greatly. This suggests a sensitivity to the individual masses, and not just the cumulative mass. As discussed in §7.4 of \citet[][]{binneyandtremaine2008}, the sensitivity of heating (the growth of random velocities) to individual compact object's mass is expected in the continuum limit. The diffusion coefficients in the truncated Fokker-Planck equation depend on $\rho \times m$, where $\rho$ is is the mass density of the perturbers and $m$ is the mass of an individual perturbers. Therefore, more massive subhaloes are more effective at perturbing the orbits of test particles. This continuum approximation breaks down where only a few subhaloes are present.

As mentioned, the high value of $A_S$ alone is not enough to assess the exact disruption done to specific $(v_r,r)$ chevrons. Therefore, Figure~\ref{fig:many_subhalo_impact} visualises the experiment that we anticipate to be the most destructive to the chevron substructures: the case with maximum $r_{\rm peri} = 20$ kpc. From this Figure, we see that there is a very different response from the chevrons to high mass subhaloes than low mass subhaloes. The few subhaloes in the $M = 10^{10}$ M$_{\odot}$ simulation severely disrupt the energy-angle stripes and the chevrons, with $A_S = 1.047$. However, despite the non-neglible value of $A_S = 0.199$, the numerous $M=10^7$ M$_{\odot}$ subhaloes have almost no visible impact on the energy-angle stripes and the chevrons, which suggests just a damping of the amplitude of the chevrons. Once again, from Figure~\ref{fig:all_lineplot}, we note a dependence on pericentre. It is clear that when the subhaloes are confined to have larger pericentres, the value of $A_S$ is reduced.

While the ironing of chevrons below $r \approx 30$ kpc is substantial for the higher mass subhaloes, we again see the formation of multiple $(E,\theta_r)$ bunches in the bottom two panels of Figure~\ref{fig:many_subhalo_impact}, which correspond to the appearance of fragmented chevrons at radii beyond $r\approx 40$ kpc.

\subsection{Impact of Known Satellites}\label{knownsubhaloes}

\begin{figure}
    \centering
    \includegraphics[width=0.47\textwidth]{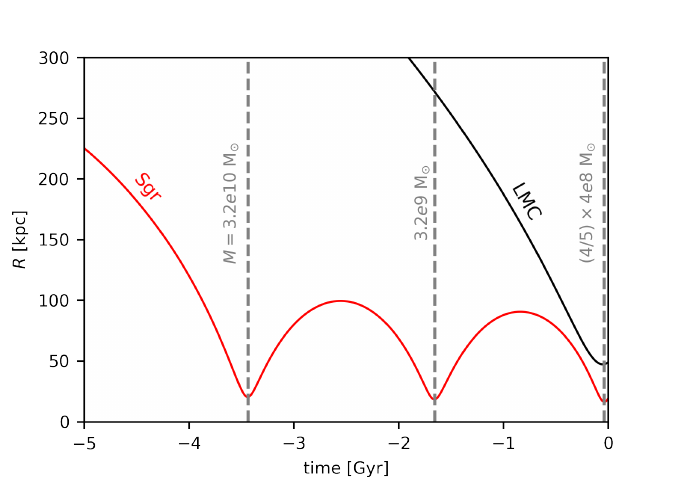}
    \caption{Galactocentric distances of the LMC and Sgr for the 5 Gyr before the present. The mass of Sgr decreases according to a decay parameter of $\delta=1.2$, and the grey text gives the mass of Sgr at each pericentre. The LMC and Sgr are assumed to be $4/5$ their known masses at $z=0$ to compensate for the reduced-mass host potential.}
    \label{fig:lmc_sgr_traj}
\end{figure}

Aside from studying the impact of generic subhaloes whose orbits are motivated only by an approximate distribution function, it is insightful to assess the impact of known massive objects whose orbits are more reasonably well understood. In this section we explore the impact of the two most relevant satellites for our study: the LMC and Sgr. They are relevant in that they have either sufficiently high mass, or sufficiently low pericentres, within the last several Gyr. Should the $(v_r,r)$ chevrons have survived the time between the last major merger and the present, it is useful to know whether the chevrons should have been partly or entirely smoothed by recent action of these two massive satellites. Moreover, knowing the expected impact on the chevrons from these satellites, it may be possible to constrain the satellites' properties using chevrons found observationally.

Since the LMC has only just reached its first pericentre, we can assume its gravitational influence can be ignored in in the first 5 Gyrs of the {\it N}-body simulation described in section (\ref{simingthesausage}). The potential of the host is again the multipole expansion potential, produced from the last snapshot {\it N}-body simulation (at a look-back time of 5 Gyr). However, since this host potential is known to be only 4/5 the mass of the MW (specifically the \textsc{MilkyWayPotential} from \textsc{Gala}), we attempt to account for this difference in by simply rescaling the mass of the LMC to be 4/5 it's value in \citet{erkal2019total}, giving $M_{\rm LMC} = 1.1 \times10^{11}$ M$_{\odot}$.

Unlike the LMC, Sgr is significantly lighter at $z=0$, with a mass of $M_{\rm Sgr}\simeq{4\times10^8}$ M$_{\odot}$ \citep[][]{vasiliev2020last}. Therefore, adjusted for the reduced mass host, we take $M_{\rm Sgr} = 3.2\times10^8$ M$_{\odot}$. However, since we compensate for mass during accretion via the same method as \citet[][]{dillamore2022impact}, Sgr could have a much heavier mass only a few pericentres ago. Additionally, Sgr's pericentre is far less than the LMC, approximately 16 kpc, and may have still been sufficiently small back when it's mass was almost $10^{11}$ M$_{\odot}$. All this in mind, Sgr may well have had a significant disruptive impact on some chevrons with peaks greater than 16 kpc.

Figure~\ref{fig:lmc_sgr_traj} illustrates the impact of a) the LMC alone and b) the LMC \& Sgr on the satellite debris. The LMC has an ironing value of $A_S = 0.022$, whereas Sgr has an value of $A_S=0.701$. The fact that Sgr results in a far greater ironing parameter than the LMC gives us an immediate expectation of their relative importance in distubring the chevrons. The first visual thing to note is the existence of chevrons at $r\sim20$--$30$ in the case of the LMC, which are far less clear when Sgr is also present (marked in red in both Figures). Additionally, note the ironing of the more central chevrons around $r \sim 25$ kpc (marked in red in both Figures) when Sgr is introduced. In this scenario, where the value of the mass decay parameter is set to be $\delta = 1.0$, the pericentre of Sgr is $\sim20$ kpc when it's mass is $3.2\times10^{10}$ M$_{\odot}$. As a result, the orbit of Sgr makes the chevrons with peaks beyond $r>20$ far more smooth than in the case without Sgr. Obviously this result is contingent upon the numerous assumptions we made about Sgr's orbit, yet it still shows that high mass objects with sufficiently low pericentres can cause chevrons to smooth out in a more sophisticated scenario where the reflex motion is accounted for. With this in mind, it may be possible to eventually constrain the mass or orbit of Sgr, should sufficient data out beyond $r>16$ kpc be collected. However, despite the LMC's large mass, with a pericentre at $r\sim50$ kpc it seems unlikely that any significant claims could be made.

\begin{figure}
    \centering
    \includegraphics[width=0.47\textwidth]{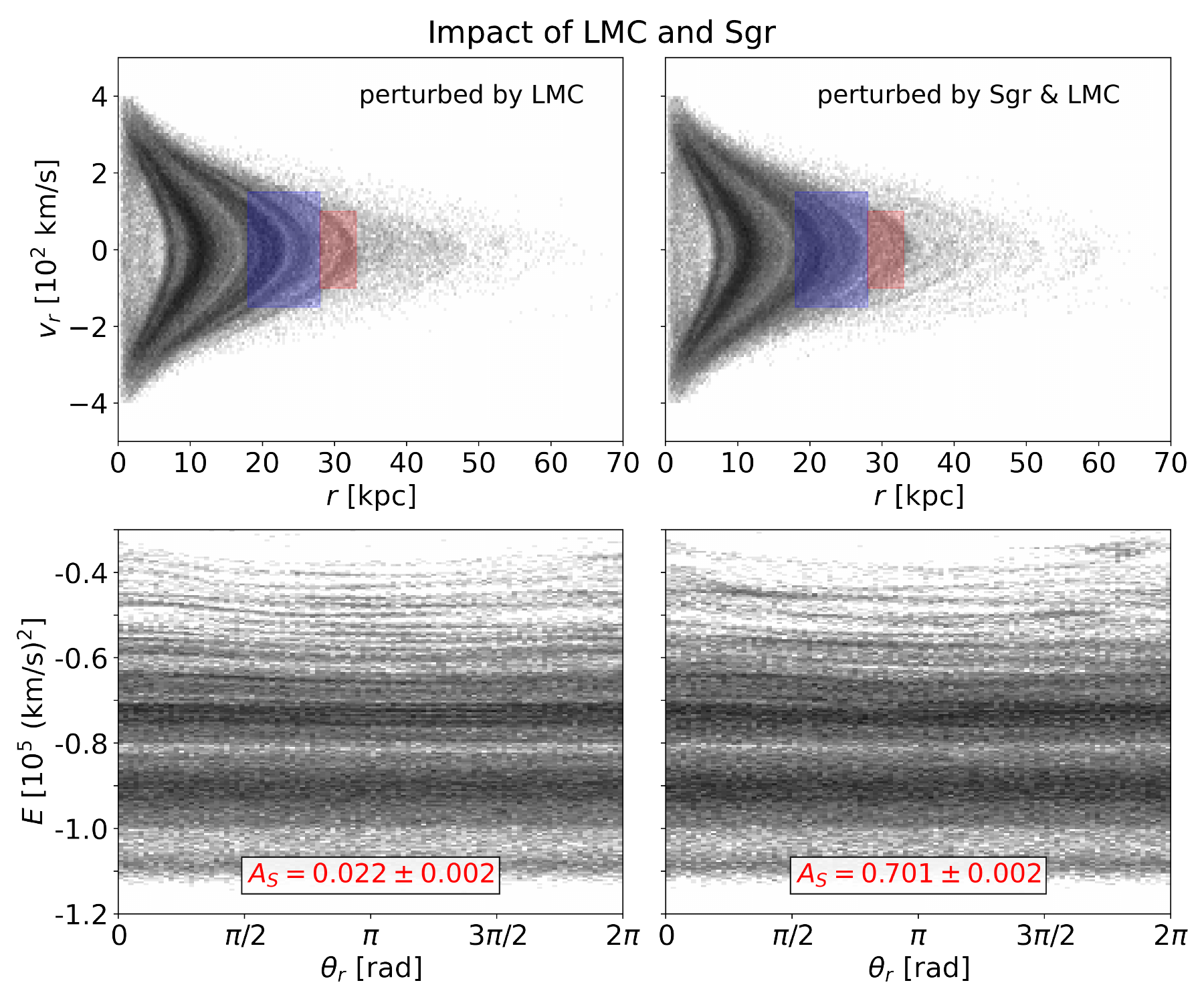}
    \caption{Impact of the LMC and Sgr on the satellite debris in $(v_r, r)$ and $(E, \theta_r)$ coordinates. From left to right rows: the unperturbed case (for comparison), the perturbation from the LMC alone, and the perturbation from the LMC and Sgr. The units of $A_S$ are (km/s)$^{-2}$rad$^{-1}$. {\it Top row:} $(v_r, r)$ space. Note the absence of the some chevrons in the right panel which are present in the left panel due Sgr larger mass several pericentres ago. {\it Bottom row:} $(E, \theta_r)$ space, with values $A_S$ calculated and shown in red text. The blue and red shaded regions mark out some chevrons which are only absent when Sgr is included.}
    \label{fig:lmc_sgr_impact}
\end{figure} 

\section{Conclusions}\label{conclusions}

In this work, we introduce the concept of using the finely substructured phase space of an ancient high mass-ratio merger to detect subhaloes. A sufficiently large and eccentric merging satellite will spew its debris across a wide range of radii of the host. The debris then phase mixes and forms wrapping chevrons in radial velocity versus radius or $(v_r,r)$ space, which match to stripes in energy versus radial angle or $(E, \theta_r)$ space that are much more visually simple than the phase space chevrons. These chevrons could be utilised in a similar fashion to cold stellar streams in constraining the mass and orbital properties of perturbing subhaloes \citep[first discussed by][]{ibata2002uncovering}. Newly discovered phase space substructure in the local stellar halo \citep[][]{belokurov2022energy} marks the transformation of this method from a theoretical exercise to a practical reality.

To investigate this idea, we explore the impact of perturbing dark matter subhaloes on the $(v_r,r)$ and $(E, \theta_r)$ substructure from phase-mixed satellite debris resulting from a large merger akin to the {\it Gaia} Sausage/Enceladus. To quantitatively show the sensitivity of the chevrons to perturbers in our simulations, we define a new quantity which we label the \textit{ironing} parameter, $A_S$, which makes use of the visual simplicity of the $(E,\theta_r)$ coordinates. To calculate $A_S$, we compare the energy-angle distribution of the unperturbed debris with the subhalo perturbed debris. Specifically, we utilise the unique nature of the $(E, \theta_r)$ substructure -- a series of thin, approximately horizontal -- and Fourier transform the energy-angle coordinates into energy and angle frequencies $(\nu_E, \nu_{\theta})$ for both the unperturbed and perturbed case. From this, we calculate the power spectra for both scenarios and subtract them to find the find the positive excess of the difference. In $(v_r, r)$ space, the phase-mixing manifests as a series of chevrons which correspond to the energy-angle stripes. Both the stripes and chevrons are smoothed out when disturbed by a sufficiently massive subhalo. While the ironing parameter's utility is currently limited to simulations, it provides a clean insight into the sensitivity of the chevrons to perturbers with a range of masses.

We conducted a series of experiments to investigate how the impact done to the phase space chevrons depends on the number of subhaloes, their mass, pericentre, and number of fly-bys through the debris. First we considered the effect of single subhaloes one at a time. In this case, the masses of the subhaloes -- represented as Hernquist potentials sat atop test particles with pre-determined orbits -- ranged from $M = 10^6$ M$_{\odot}$ up to $10^{10}$ M$_{\odot}$. Additionally, the orbits of the subhaloes were set up with a variety of pericentres: $r_{\rm peri} = 10, 30$ and $50$ kpc. Chevrons whose maximum radial extent ($r_s$ in equation \ref{eq:chev_quadratic}) was approximately greater than that of the subhalo pericentre are smoothed out, provided the subhalo has sufficiently high mass; ironing of chevrons is only significant for the $10^{10}$ M$_{\odot}$ case. We considered the impact of single subhaloes after one fly-by, calculating the value of $A_S$ after one orbital period, and compared this with the value of $A_S$ after multiple fly-bys. In all cases, multiple fly-bys increased the value of $A_S$. However, we found that, no matter the length of time of integration, subhaloes with $M\leq 10^8$ M$_{\odot}$ had a mostly negligible impact on the phase space.

Second, we considered the effect of multiple subhaloes, whose mass-number combination was determined by the subhalo mass function i.e. equations (\ref{eq:SHMF}) and, (\ref{eq:integrateshmf}). The position and velocity of these subhaloes was sampled from an appropriate distribution function approximately the distribution function of known Milky Way satellites. Here, we considered 4 subhaloes of mass $10^{10}$ M$_{\odot}$, 28 subhaloes of mass $10^{9}$ M$_{\odot}$, 225 subhaloes of mass $10^{8}$ M$_{\odot}$ and 1789 subhaloes of mass $10^{7}$ M$_{\odot}$. We considered two sub-cases: a sampling of subhaloes such that their pericentres were forced to have maximum value of 20 kpc, and 50 kpc. From this experiment we learned that the phase space chevrons were likely sensitive to individual subhalo mass, and not just cumulative perturbing mass, with the value of $A_S$ increasing as the mass increased, for both pericentres. Moreover, in all cases visualised, there remained some local ($r \lesssim 20$ kpc) substructure in $(v_r, r)$ space, adding to the confidence in the discovery by \citet[][]{belokurov2022energy}. 

Given the relationship between chevron impact on perturber mass and pericentre, it is theoretically possible to constrain the orbit of a large mass subhalo provided sufficient $(v_r, r)$ data exists. Therefore, the existence of chevrons with tips in the $r = 20$--$30$ kpc region, in observational data, may be a tool for constraining the mass or orbit of Sgr. Moreover, the chevrons detected in the local stellar halo by \citet[][]{belokurov2022energy} could provide a new method of constraining the subhalo mass function, and therefore the $\Lambda$CDM model of galaxy formation, as the visibility of the chevrons depends on the number and mass of subhaloes present in the Milky Way. We intend to follow up this work with a more detailed investigation into the effects of perturbers on individual chevrons, expanding on what is shown in Figure~\ref{fig:individual}. It may be useful to more thoroughly examine the dependence of chevron sensitivity to chevron energy, peak radius and other properties.

However, subhaloes are not the only source of perturbation that could disturb the phase space chevrons. It is now well known that the Milky Way has a central bar of approximately 3 to 5 kpc across \citep[e.g.][]{lucey2022constraining}. In a future paper we intend to explore the impact of a central rotating bar on the chevrons, particularly at low galactocentric radii. 

\appendix

\section*{Acknowledgements}

We thank the anonymous referee for helpful comments. EYD thanks the Science and Technology Facilities Council (STFC)
for a PhD studentship (UKRI grant number 2605433), and is grateful to the Center for Computational Astrophysics (CCA) for hospitality during his stay, where some of this work was completed. AMD thanks STFC for a PhD studentship (UKRI grant number 2604986).  For the purpose of open access, the author has applied a Creative Commons Attribution (CC BY) license to any author accepted manuscript version arising.

\section*{Data Availability}

The simulations in this project can be reproduced with publicly available software, using the description provided in Section~\ref{simulations}.



\bibliographystyle{mnras}
\bibliography{ironing} 




\appendix

\section{Tables}

In this appendix we present the tables of the calculated values of $A_S$ and $\sigma_A$ from the experiments conducted in Section~\ref{genericsubhaloes}.

\begin{table}
\centering
\caption{Values of the ironing parameter $A_S$ for single subhalo, single fly-by experiments.}
\resizebox{0.48\textwidth}{!}{%
\begin{tabular}{|l|lll|}
\hline
\multirow{3}{*}{Subhalo Mass [M$_{\odot}$]} & \multicolumn{3}{c|}{$A_S$ [(km/s)$^{-2}$rad$^{-1}$]} \\ \cline{2-4}
 & \multicolumn{3}{c|}{$r_{\rm peri}$ [kpc]} \\ 
 & \multicolumn{1}{c|}{10} & \multicolumn{1}{c|}{30} & \multicolumn{1}{c|}{50} \\ \hline
$10^{10}$ & \multicolumn{1}{l|}{$(0.723 \pm 0.002)$} & \multicolumn{1}{l|}{$(0.543\pm 0.002)$} & $(0.309\pm 0.002)$ \\ \hline
$10^{9}$ & \multicolumn{1}{l|}{$0.239$} & \multicolumn{1}{l|}{$0.125$} & $0.063$ \\ \hline
$10^{8}$ & \multicolumn{1}{l|}{$0.032$} & \multicolumn{1}{l|}{$0.018$} & $0.010$ \\ \hline
$10^{7}$ & \multicolumn{1}{l|}{$0.010$} & \multicolumn{1}{l|}{$0.004$} & $0.003$ \\ \hline
$10^{6}$ & \multicolumn{1}{l|}{$0.003$} & \multicolumn{1}{l|}{$0.002$} & $0.001$ \\ \hline
\end{tabular}%
}
\label{tab:n1_area_singleflyby_table}
\end{table}

\begin{table}
\centering
\caption{Values of the ironing parameter $A_S$ for single subhalo, many fly-by experiments.}
\resizebox{0.48\textwidth}{!}{%
\begin{tabular}{|l|lll|}
\hline
\multirow{3}{*}{Subhalo Mass [M$_{\odot}$]} & \multicolumn{3}{c|}{$A_S$ [(km/s)$^{-2}$rad$^{-1}$]} \\ \cline{2-4} 
 & \multicolumn{3}{c|}{$r_{\rm peri}$ [kpc]} \\  
 & \multicolumn{1}{c|}{10} & \multicolumn{1}{c|}{30} & \multicolumn{1}{c|}{50} \\ \hline
$10^{10}$ & \multicolumn{1}{l|}{$(0.889 \pm 0.002)$} & \multicolumn{1}{l|}{$(0.655 \pm 0.002)$} & $(0.336 \pm 0.002)$ \\ \hline
$10^{9}$ & \multicolumn{1}{l|}{$0.310$} & \multicolumn{1}{l|}{$0.189$} & $0.079$ \\ \hline
$10^{8}$ & \multicolumn{1}{l|}{$0.057$} & \multicolumn{1}{l|}{$0.025$} & $0.014$ \\ \hline
$10^{7}$ & \multicolumn{1}{l|}{$0.014$} & \multicolumn{1}{l|}{$0.007$} & $0.003$ \\ \hline
$10^{6}$ & \multicolumn{1}{l|}{$0.005$} & \multicolumn{1}{l|}{$0.002$} & $0.002$ \\ \hline
\end{tabular}%
}
\label{tab:n1_area_multiflyby_table}
\end{table}

\begin{table}
\centering
\caption{Values of the ironing parameter $A_S$ for multiple subhaloes, many fly-by experiments.}
\resizebox{0.4\textwidth}{!}{%
\begin{tabular}{|l|ll|}
\hline
\multirow{2}{*}{Subhalo Mass [M$_{\odot}$]} & \multicolumn{2}{c|}{$A_S$ [(km/s)$^{-2}$rad$^{-1}$]} \\ \cline{2-3} 
 & \multicolumn{2}{c|}{max $r_{\rm peri}$ [kpc]} \\  
 & \multicolumn{1}{c|}{20} & \multicolumn{1}{c|}{50} \\ \hline
$10^{10}$ & \multicolumn{1}{l|}{$(1.047\pm 0.002)$}  & $(0.847\pm 0.002)$ \\ \hline
$10^{9}$ & \multicolumn{1}{l|}{$0.717$}  & $0.639$ \\ \hline
$10^{8}$ & \multicolumn{1}{l|}{$0.423$}  & $0.399$ \\ \hline
$10^{7}$ & \multicolumn{1}{l|}{$0.199$}  & $0.134$ \\ \hline
\end{tabular}%
}
\label{tab:nmany_area_multiflyby_table}
\end{table}

\section{Subhalo Concentration}

\begin{figure}
    \centering
    \includegraphics[width=0.43\textwidth]{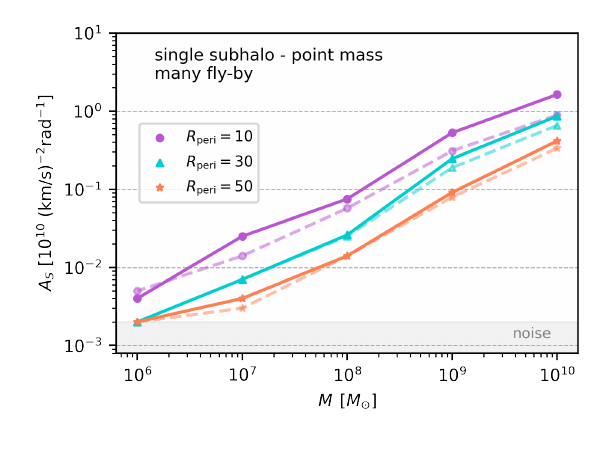}
    \caption{The value of the ironing parameter $A_S$ for single subhalo, many flyby experiment, where we have replaced the subhaloes' Hernquist potentials with a point mass potential. The faded dashed lines show the trend for Hernquist potential whereas the solid line shows the trend for the point mass potential. While the values are slighly different, we see that the approximate power law trend is preserved in the case of the point mass potentials.}
    \label{fig:pointmass_lineplot}
\end{figure}

To assess the robustness of the approximate power law trend between $A_S$ and mass, we reconduct experiment (ii) but replace the Hernquist potentials with a point mass potential i.e. a Plummer sphere with a scale radius equal to zero. The results of this additional experiment are shown as a bold line in Figure~\ref{fig:pointmass_lineplot}, alongside the results of experiment (ii) in a dashed faded line. We see that the approximate power law trend is preserved for the new values of $A_S$. Note that for more concentrated subhaloes the value of $A_S$, and therefore the impact on the chevrons, is increased. A more concentrated subhalo encases a larger amount of mass within the same radius than a less concentrated halo. Hence the force on the particles is larger at all radii.

\bsp	
\label{lastpage}
\end{document}